\documentclass[apj]{emulateapj}
\usepackage{longtable}
\newcommand{\be}{\begin{equation} }
\newcommand{\ee}{\end{equation} }
\newcommand{\bea}{\begin{eqnarray} }
\newcommand{\eea}{\end{eqnarray} }

\newcommand{\teff}{${\rm T_{eff}}$ }
\slugcomment{Accepted to ApJ: 15 July 2007}
\shortauthors{Maness {\it et~al.\/}}
\shorttitle{GC Top-Heavy IMF}
\begin{document}
\title{~~\\ ~~\\ Evidence for a Long-Standing Top-Heavy IMF 
in the Central Parsec of the Galaxy}  
\author{ H. Maness\altaffilmark{1},
F. Martins\altaffilmark{2}, S. Trippe\altaffilmark{2},
R. Genzel\altaffilmark{2, 3}, J. R. Graham\altaffilmark{1},
C. Sheehy\altaffilmark{4}, M. Salaris\altaffilmark{5},
S. Gillessen\altaffilmark{2},
T. Alexander\altaffilmark{6}, T. Paumard\altaffilmark{2},
T. Ott\altaffilmark{2}, R. Abuter\altaffilmark{2},
F. Eisenhauer\altaffilmark{2} } 
\altaffiltext{1}{Department of
Astronomy, University of California at Berkeley, Berkeley, USA}
\altaffiltext{2}{Max-Planck Institut f${\rm \ddot{u}}$r
extraterrestrische Physik (MPE), Garching, FRG}
\altaffiltext{3}{Department of Physics, University of California at
Berkeley, Berkeley, USA} \altaffiltext{4}{Department of Astronomy,
University of Chicago, Chicago, USA} \altaffiltext{5}{Astrophysics
Research Institute, Liverpool John Moores University, Liverpool, UK}
\altaffiltext{6}{Faculty of Physics, Weizmann Institute of
Science, Rehovot, Israel}

\begin{abstract}
We classify 329 late-type giants within 1 parsec of Sgr A$^{*}$, using
the adaptive optics integral field spectrometer SINFONI on the VLT.
These observations represent the deepest spectroscopic data set so far
obtained for the Galactic Center, reaching a 50\% completeness
threshold at the approximate magnitude of the helium-burning red clump
($K_{S} \sim 15.5$ mag.).  Combining our spectroscopic results with
NaCo $H$ and $K_{S}$ photometry, we construct an observed
Hertzsprung-Russell diagram, which we quantitatively compare to
theoretical distributions of various star formation histories of the
inner Galaxy, using a $\chi^2$ analysis.  Our best-fit model
corresponds to continuous star formation over the last 12 Gyr with a
top-heavy initial mass function (IMF).  The similarity of this IMF to
the IMF observed for the most recent epoch of star formation is
intriguing and perhaps suggests a connection between recent star
formation and the stars formed throughout the history of the Galactic
Center.
\end{abstract}


\section{Introduction}

A growing body of evidence suggests that the stellar population of the
Galactic nucleus is distinct from that of the bulge.  Surface brightness
measurements from the NIR to the FIR show that in contrast to the
bulge, the central few hundred parsecs of the Galaxy are dominated by
a flat, disk-like distribution of stars, gas, and dust
\citep{Kent92, Launhardt02}.  This difference is also reflected in the
kinematics: OH/IR stars in the central hundred parsecs show higher
rotational velocities than expected for an inner single-component,
bulge population \citep{Lindqvist92}.

Evidence for ongoing star formation also distinguishes the nucleus
from the bulge.  The Galactic bulge is composed primarily of an old
starburst-like population formed early, some 7--10 Gyr ago
\citep{Zoccali03,vanLoon03,Zoccali06,Ballero07}.  In contrast, the
Galactic nucleus harbors substantial young and intermediate-age
stellar populations.  Intermediate-age populations have been most
commonly inferred from broad-band photometry \citep{Rieke87,
Narayanan96, Davidge97, Philipp99, Alexander99, Figer04}.  Star
formation tracers have also been used to study intermediate-age
populations, including young supergiants and luminous AGBs, which
trace 10 Myr - 1 Gyr populations \citep{Sellgren87, Blum96b, Blum96a},
and OH/IR stars, which trace 1-3 Gyr populations \citep{Wood98,
Sjouwerman99}.  The young stellar populations in the Galactic nucleus
have been the subject of much recent work.  These populations are
predominantly concentrated in three young, massive clusters: the
Arches and Quintuplet clusters (Figer et al. 1999, 2002; Stolte et
al. 2005, and references therein), approximately 30-50 pc in projected
distance from the Galactic Center, and the Central Cluster, located
within the central parsec (Krabbe et al. 1991, 1995; Paumard et
al. 2006, and references therein).

Star formation in the nucleus is thought to be ultimately linked to
the inward transport of gas induced by the Galactic bar
\citep{Morris96,Kormendy04}.  However, precisely how gas is funneled
from the outer galaxy to the nucleus and how this process affects the
resultant star formation, remains poorly known.  At present, there is
a concentration of molecular gas 2-8 pc from the Galctic Center termed
the circumnuclear disk \citep{Guesten87, Jackson93, Serabyn94}.  If
such a structure routinely fuels star formation in the central parsec,
the stars formed in this region may represent an entirely distinct
population from the Arches and Quintpulet clusters or the intermediate
age populations found throughout the $\sim$200 pc nucleus.  Dynamical
effects due to the supermassive black hole (Genzel et al. 2003, Ghez
et al. 2005, and references therein) and the large stellar and remnant
density in the central parsec may also lead to population differences
between the Central Cluster and that in the larger nucleus, either
prior to star formation or afterwards.

A number of investigators have attempted to address the nature of the
Central Cluster and its relationship to the larger 100-300 pc nucleus.
Most recently, \citet{Paumard06} investigated the properties of the
young stellar population in the central parsec, spectroscopically
identifying nearly 100 OB and Wolf-Rayet stars.  They find that the
majority of the young stars reside in two, inclined and
counter-rotating disks, suggesting in situ star formation in dense gas
accretion disks.  Studies of the older stellar population in the
central parsec have been primarily limited to broadband photometry.
They have suggested that the fraction of low mass stars in the
population increases with distance from the center, perhaps due to
dynamical mass segregation \citep{Philipp99,Genzel03,Schodel07}.

A detailed analysis, however, of the late-type giant population
requires spectroscopy, owing to the large scatter in extinction near
the Galactic Center and the intrinsic variations in giant star colors.
\citet{Blum03} pioneered work in this area, using spectroscopic and
photometric observations of the most luminous giants and supergiants
(50\% complete at $K_{S}\sim$10 mag.) to construct an H-R diagram for
stars in the inner 5 pc.  They report that the GC star formation rate
in the central few parsecs is largely similar to that of the bulge.
Specifically, they find that the majority of stars formed more than 5
Gyr ago, though they also find evidence that significant star
formation also occurred during the past 100 Myr.  However, their
conclusions are limited by their bright magnitude limit, which samples
only short-lived evolutionary stages, for which theoretical models are
uncertain.

In this paper, we build on the work of \citet{Blum03} in an effort to
better characterize the late-type giant population within the central
parsec.  We report deep photometric and spectroscopic observations of
329 late-type giants in the GC, complete to 50\% at $K_{S}\sim$15.5
mag.  Our observations for the first time include the helium-burning
red clump, as well as the red giant branch and asymptotic giant
branch.  This improved magnitude limit allows for the most robust
picture of the Galactic Center star formation history to date, as
these stars are much better understood than the supergiants and
luminous AGB stars studied by \citet{Blum03}.  In \S 2, we present our
observations and the techniques used to construct the
Hertzsprung-Russell (H-R) diagram.  Section 3 discusses the resulting
H-R diagram and implications for the GC star-formation history.  We
discuss our results in \S 4 and conclude with \S 5.

\section{Observations and Data Reduction}

\subsection{Spectroscopic Observations and \teff Determination}

We observed eight Galactc Center fields between March and September
2006 using the integral field spectrometer SPIFFI
\citep{Eisenhauer03a,Eisenhauer03b} in conjunction with the MACAO
adaptive optics module \citep{Bonnet03} mounted on the SINFONI ESO VLT
facility.  We observed these fields for the dual purpose of
identifiying main sequence B stars outside the central parsec (Martins
et al. 2007b, in prep) and gathering spectroscopic observations for a
large number of late-type giants.  We chose fields with a radial
distance from Sgr A$^*$ of less than $20$ arcsec, selecting regions
outside the minispiral to avoid nebular contamination in Brackett
$\gamma$, and choosing fields north of Sgr A$^*$ to minimize
separation from the AO guide star. We also avoided regions with
extremely bright stars ($K \lesssim 9$).  As stars this bright are
relatively rare, the selection bias in avoiding these stars is
negligible.  If the fields had instead been chosen at random, we
expect no more than a few $K \lesssim 9$ stars would be selected.

\begin{figure}
\epsscale{1.0}
\plotone{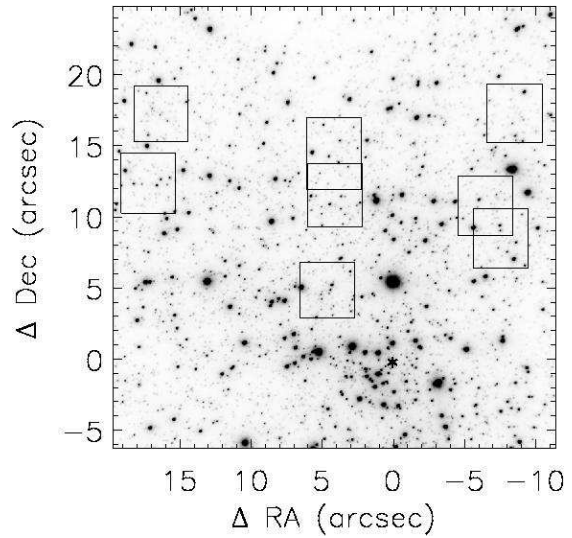}
\caption{NaCo $K_{S}$ band mosaic used to derive photometry.  
The eight regions for which we have SINFONI spectra are overplotted 
in black.  The position of Sgr A$^{*}$ is shown with an $*$.}
\label{mos}
\end{figure}

\begin{figure}
\epsscale{0.7}
\plotone{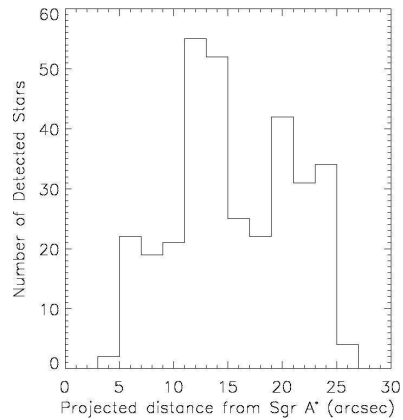}
\caption{Histogram of stellar positions, corresponding to the stars
detected in the spectroscopic fields marked in Figure \ref{mos}.}
\label{dist_hist}
\end{figure}

We observed each selected SINFONI field in 50 $\times$ 100 mas pixel
mode, resulting in 0.2 arcsec resolution and a $4.2'' \times 4.2 ''$
field of view.  We simultaneously observed all fields in $H$ and $K$
band, leading to a spectral resolution of R $\approx$ 1500. Figure
\ref{mos} displays the location of the observed fields.  The total
exposure time for all fields was 4200 s, except for the fields at
(17$''$, 17$''$), (5$''$, 11$''$), and (-8$''$, 8$''$), which had
total exposure times of 6600 s, 7200 s, and 2400 s, respectively.  The
fields span a range in projected distance from Sgr A* of 4 arcsec to
26 arcsec.  A histogram of the observed stellar positions is shown in
Figure \ref{dist_hist}.

\begin{figure*}
\epsscale{0.6}
\includegraphics[width=.7\textwidth,angle=90]{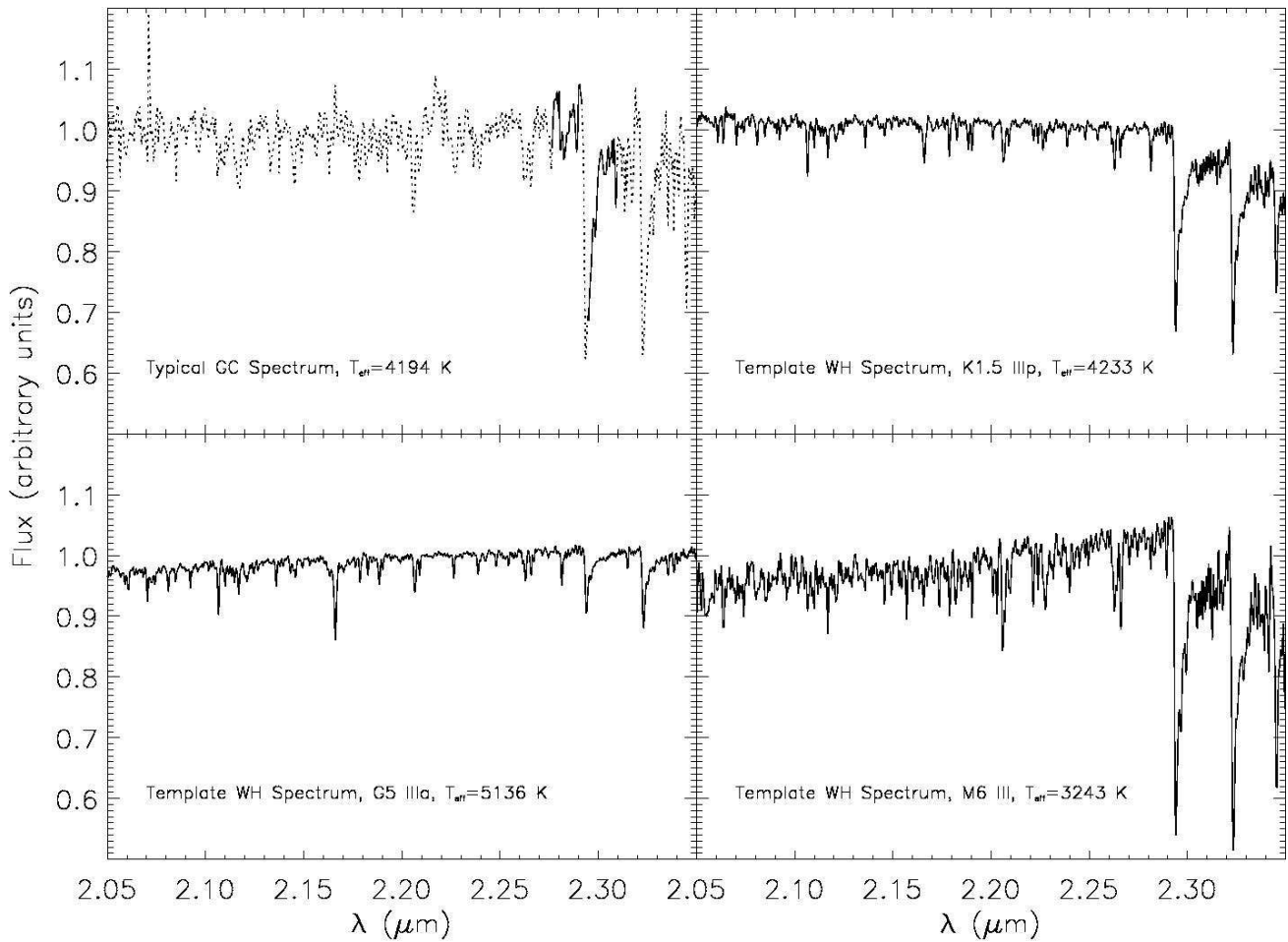}
\caption{A typical stellar spectrum from the Galactic Center fields.
The position of the $^{12}$CO 2.2935 $\mu$m $\nu$=2-0 rovibrational
bandhead and nearby continuum used to calculate the CO-index is shown
as a solid line.  Three template spectra from the \citet{Wallace97}
spectral library are also shown for comparison.}
\label{spectrum}
\end{figure*}

We reduced the raw data using a standard procedure to perform
flat-fielding, sky subtraction, and wavelength calibration
\citep{Schreiber04}.  Following extraction, we removed the stellar
continua by dividing by a second degree polynomial fitted to the
line-free regions of the stellar spectra.  The normalized spectra
allow us to directly compare the VLT spectra to the normalized catalog
spectra in \citet{Wallace97} and \citet{Kleinmann86}.  A typical
resultant GC $K$ band spectrum following normalization is shown in
Figure \ref{spectrum}.  Three template spectra from the
\citet{Wallace97} spectral library are also shown for comparison.

To classify the stars in our sample, we used the $^{12}$CO 2.2935
$\mu$m, $\nu=2-0$ rovibrational band head, the strongest feature in
our spectra and a well-known \teff indicator, for a given luminosity
class \citep{Kleinmann86}.  We use only this feature, as it is largely
free of nebular and telluric lines, making it more reliable than other
$H-$ and $K-$band signatures in our spectra.  We note that in contrast
to the work of \citet{Blum96b, Blum03}, we do not need the H+K band
H$_2$O feature to reliably predict effective temperatures, as our
sample consists only of giants with $K_S>$10.3 (see \S2.2).  Due to
their rarity, stars sufficiently bright to be supergiant or long
period variable candidates are not contained in our sample.

\begin{figure}
\epsscale{0.2}
\includegraphics[width=.4\textwidth,angle=90]{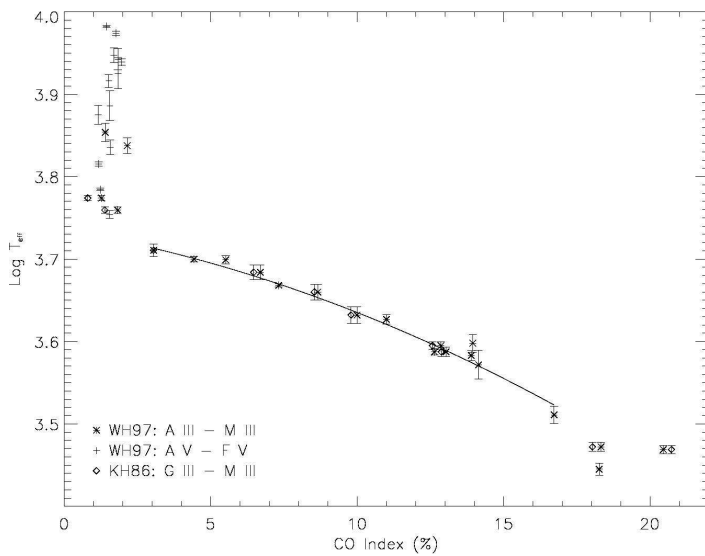}
\caption{The 2.3 $\mu$m CO index (see text for definition) as a
function of temperature for the comparison stars taken from the 
\citet{Wallace97} and \citet{Kleinmann86} catalogs.  The solid line is a 
least-squares fit to both data sets for stars with  
17 $>$ CO $\ge$ 3.}
\label{comp_relation}
\end{figure}

To define a \teff- $^{12}$CO index relation, we computed the CO index
defined by \citet{Blum96b}. Although not as well-known as other CO
indices (e.g. equivalent width, photometric indices of
\citet{Frogel78}), \citet{Blum96b} showed that this index correlates
well with other CO indices.  Furthermore, it has the advantage of
being insensitive to small variations in the nearby continuum because
it does not require a fit to the continuum.  The index is defined as
CO \% = [(1 $-$ F${\rm _{band} /}$ F${\rm _{cont}}$) $\times$ 100)],
where both bands are 0.015 $\mu$m wide and the continuum and CO band
centers are defined as 2.284 $\mu$m and 2.302 $\mu$m, respectively.
We note that although we adopt the same definition for the CO index as
\citet{Blum96b}, our spectra are normalized while theirs are not, and
thus, our computed indices are not directly comparable to theirs.  We
calculated uncertainties in the computed indices, assuming the noise
is dominated by photon statistics of the source and background and
that the uncertainty in the band is approximately equal to that of the
nearby continuum.

Following \citet{Blum03}, we used CO strength to estimate effective
temperatures for the cool giants in our GC sample.  To define a CO
index versus \teff relation, we used a set of archival comparison star
spectra taken from \citet{Wallace97} and \citet{Kleinmann86} with
well-determined effective temperatures given in the literature.  In an
effort to avoid systematic errors in our derived index$-$\teff
relation, we used a large number of references, assuming temperatures
derived from a wide variety of methods.  A summary of the comparison
star data is given in Table \ref{comp_table}, and the resulting
relationship between CO index and \teff is shown in Figure
\ref{comp_relation}.  The \citet{Wallace97} and \citet{Kleinmann86}
relationships are in very good agreement with each other (for stars in
common, $\Delta$CO $< 0.5 \%$), suggesting that the CO index is
largely independent of resolution or observing system.  The relation
is tightest for CO $\lesssim 17$.  As no stars in our GC sample have
computed indicies exceding this value, we did not include comparison
stars with CO $> 17$ in our fit.  The resulting index-\teff relation
using both empirical data sets is: $\log$ \teff = 3.7351 $-$ 0.0060
$\times$ CO $-$ 0.00040 $\times$ CO$^2$, for all stars with
measured indices $17 >$ CO $\ge$ 3.  The errors in the derived
coefficients are 0.0092, 0.0022, and 1.2$\times10^{-4}$, respectively.

\begin{deluxetable*}{llllllll}
\tablecaption{Comparison Star Properties}
\tablecolumns{8}
\tablewidth{0pt}
\tablehead{
\colhead{Name} & \colhead{Sp Type} & \colhead{Catalog} & 
\colhead{CO} & \colhead{$\Delta$CO} &
\colhead{\teff (K)} & \colhead{$\Delta$\teff}
& \colhead{\teff Reference} \\
 & & &  (\%) &  (\%) & (K) & (K) &
}
\startdata
HR 7001	 & A0 Va        & WH       & 1.760  &  0.004 & 9420 &  60    &  \citet{Smalley93}	\\
HR 4295	 & A1 V         & WH       & 1.434  &  0.003 & 9600 &  25    &  \citet{Adelman02}	\\
HR 5054	 & A1 Vp        & WH       & 1.823  &  0.004 & 8760 &  260   &  \citet{Sokolov98}	\\
HR 6378	 & A2 V         & WH       & 1.953  &  0.004 & 8690 &  78    &  \citet{Blackwell98} \\
HR 4534	 & A3 V         & WH       & 1.695  &  0.005 & 8857 &  185   &  \citet{Malagnini97} \\
HR 4357	 & A4 V         & WH       & 1.516  &  0.007 & 8243 &  150   &  \citet{Smalley93}	\\
HR 7557	 & A7 V         & WH       & 1.147  &  0.004 & 7500 &  200   &  \citet{Theodossiou91} \\
HR 2943	 & F5 IV-V      & WH       & 1.176  &  0.014 & 6532 &  39    &  \citet{Ramirez05}	\\
HR 6927	 & F7 V         & WH       & 1.219  &  0.014 & 6087 &  22    &  \citet{Taylor03}	\\
HR 4375	 & F8.5 V       & WH       & 1.540  &  0.015 & 5676 &  66    &  \citet{Taylor03}	\\
HR 21	 & F2 III       & WH       & 1.563  &  0.011 & 6847 &  137   &  \citet{Blackwell94}	\\
HR 403	 & A5 III-IV    & WH       & 1.828  &  0.008 & 8420 &  360   &  \citet{Malagnini90}	\\
HR 1412	 & A7 III       & WH       & 1.554  &  0.007 & 7690 &  320   &  \citet{Sokolov98}	\\
HR 1457	 & K5$+$ III    & WH       & 12.633 &  0.023 & 3866 &  35    &  \citet{Alonso99}      \\
HR 2985	 & G8 III       & WH       & 5.540  &  0.018 & 5001 &  56    &  \citet{Alonso99}      \\  
HR 3003	 & K4 III       & WH       & 13.944 &  0.023 & 3961 &  99    &  \citet{Alonso99}	\\           
HR 3323	 & G5 IIIa      & WH       & 3.045  &  0.017 & 5136 &  88    &  \citet{Alonso99}	\\     
	 &		& KH	   & 3.039  &  0.007 & & 			\\
HR 4031	 & F0 III       & WH       & 2.145  &  0.011 & 6880 &  150   &  \citet{Smalley93}	\\
HR 4069	 & M0 III       & WH       & 14.143 &  0.021 & 3730 &  149   &  \citet{Engelke92}	\\
HR 4517	 & M1 III       & WH       & 13.894 &  0.019 & 3828 &  53    &  \citet{Feast96}	\\
HR 4883	 & G0 IIIp      & WH       & 1.819  &  0.014 & 5747 &  54    &  \citet{Blackwell98}	\\
	 &		& KH	   & 1.377  &  0.007 & &			\\
HR 5017	 & F3 III       & WH       & 1.395  &  0.017 & 7141 &  181   &  \citet{Alonso99}	\\        
HR 5340	 & K1.5 IIIp    & WH       & 10.997 &  0.021 & 4233 &  55    &  \citet{Alonso99}	\\       
HR 6299	 & K2 III       & WH       & 8.653  &  0.020 & 4571 &  100   &  \citet{Bell89}	\\
	 &		& KH	   & 8.530  &  0.009 & &			\\
HR 6703	 & G8.5 III     & WH       & 4.426  &  0.020 & 5011 &  35    &  \citet{Blackwell98}	\\
HR 6705	 & K5 III       & WH       & 12.859 &  0.023 & 3934 &  42    &  \citet{Alonso99}	\\        
	 &		& KH	   & 12.569 &  0.010 & & 			\\
HR 7635	 & M0$-$ III    & WH       & 13.035 &  0.025 & 3867 &  50    &  \citet{Alonso99}	\\        
	 &		& KH	   & 12.869 &  0.010 & & 			\\
HR 7806	 & K2.5 III     & WH       & 9.993  &  0.022 & 4286 &  100   &  \citet{Bell89}	\\
	 &		& KH	   & 9.785  &  0.010 & &			\\
HR 7886	 & M6 III       & WH       & 16.722 &  0.025 & 3243 &  79    &  \citet{Perrin98}	\\
HR 8317	 & K0.5 III     & WH       & 7.319  &  0.024 & 4658 &  19    &  \citet{Gray01}	\\
HR 8694	 & K0$-$ III    & WH       & 6.691  &  0.021 & 4830 &  100   &  \citet{Bell89}	\\
	 &		& KH	   & 6.463  &  0.010 & & 			\\
HR 8905	 & F8 III       & WH       & 1.265  &  0.014 & 5942 &  42    &  \citet{Blackwell98}	\\
	 &		& KH	   & 0.797  &  0.007 & &			\\
SWVir	 & M7 III       & WH       & 18.350 &  0.024 & 2966 &  36    &  \citet{Dyck96}	\\
	 &		& KH	   & 18.033 &  0.008 & &			\\
BKVir	 & M7$-$ III    & WH       & 20.449 &  0.022 & 2944 &  34    &  \citet{Perrin98}	\\
	 &		& KH	   & 20.730 &  0.006 & &			\\
RXBoo	 & M7.5-8 III   & WH       & 18.262 &  0.024 & 2786 &  46    &  \citet{Perrin98}	\\          
\enddata
\label{comp_table}
\end{deluxetable*}

In order to separate cool giants from warmer giants and main sequence
stars, warm stars with CO $<$ 3 are removed from the sample.  Of the
355 detected stars in our spectroscopy, 329 were categorized as
CO-absorbers, and we assigned effective temperatures to these stars
using the above index$-$\teff relation.  Prior to temperature
calculation, we cross-correlated all CO-absorbing stars with a
CO-template in order to remove radial velocities.  Errors in
temperatures were estimated based on the noise in each spectrum and
the intrinsic dispersion in our index$-$\teff relation ($\sigma$ =
0.0050).  As noted by \citet{Ramirez97}, errors determined in this way
are strictly only lower limits.  The spectral classification and
analysis of the remaining early-type (non-CO-absorbing) stars in our
sample is less straight-forward \citep{Martins07a}, and we defer
analysis of these stars to a later paper (Martins et al. 2007b, in
prep).

\subsection{Photometric Observations}

We made photometric observations using the imaging system NAOS/CONICA
(NaCo), consisting of the adaptive optics system NAOS
\citep{Rousset03} and the NIR camera CONICA \citep{Hartung03} at the
8.2-m UT4 (Yepun) of the ESO VLT. In April 2006, we collected several
AO-corrected images in each of $H$ and $K_{S}$ band with a pixel size
of $\delta x =$ 27 mas.  The total exposure time was 64 s for each
band, broken into thirty-two dithered images, in which every fourth
image was separated from the previous exposure by 16 arcsec.  We used
our standard reduction pipeline to perform sky subtraction, bad pixel
corrections, flat field corrections, and stacking of images to create
final mosaics.  In Figure \ref{mos}, we display a sub-section of the
final $K_{S}$ band mosaic, containing all stars analyzed in this
paper.  Photometry was extracted for only this region to help ensure
PSF constancy across the image.

We used the crowded field photometry package StarFinder
\citep{Diolaiti00} to establish relative photometry and source
detection.  An empirical PSF for the central core (FWHM $\simeq$ 0.10
arcsec) was extracted from each image using seven bright, isolated
stars.  The full radial extent of the extracted PSF was r $\simeq$
0.19 arcsec.  To derive the photometric curve of growth needed to
place these results on an absolute scale, we adopted the MTF-fitting
technique of \citet{Sheehy06}.  This technique fits the power spectrum
of the image using a combination of the source spatial distribution
function determined by StarFinder and a parameterized description of
the modulation transfer functions (MTF) of the atmosphere, telescope,
AO system, and science camera.  The advantage of this technique is
that it derives the PSF encircled energy curve of growth, including
the extended, seeing-limited halo, and thus provides the aperture
correction required for absolute photometric calibration.  In deriving
the curve of growth from the data themselves, we avoid systematic
calibration errors introduced from using a PSF standard acquired at a
different time and under different observing conditions than the
target data.  Given the rapid variations in AO performance
\citep{Fitzgerald06, Vacca07}, this technique is crucial for deriving
accurate absolute photometry.

\begin{figure}
\epsscale{0.7}
\includegraphics[width=.35\textwidth,angle=90]{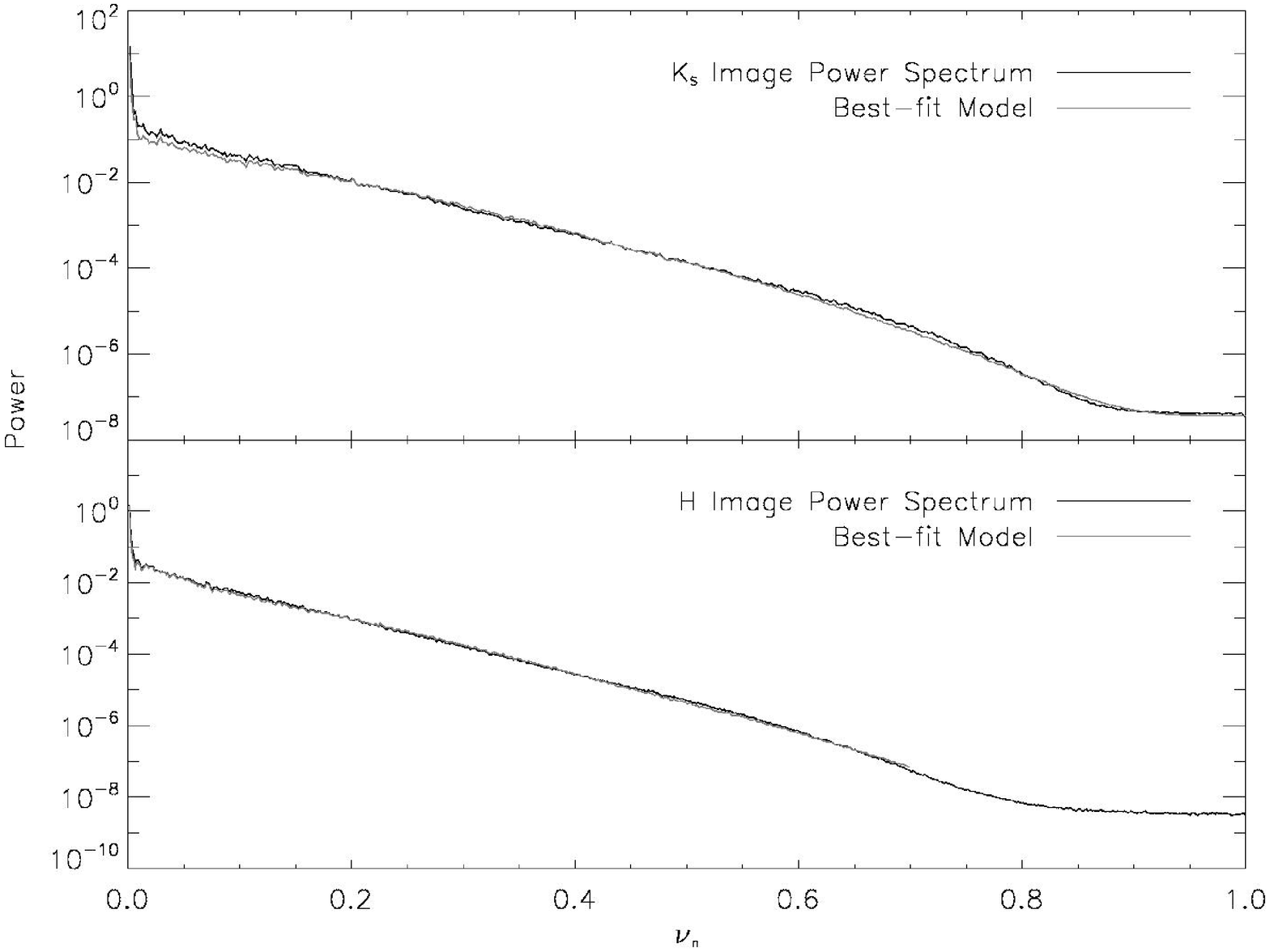}
\caption{Comparison of the image power spectra to the best-fit models
based on the parameterization of \citet{Sheehy06}.  The $H$ band data
are aliased, and we therefore, fit only the power spectrum for
$\nu_{n} < 0.7$ (see \S 2.2).  The spatial frequency, $\nu_{n}$, is
normalized relative to the cutoff frequency in the image plane 
(D$_{tel}/\lambda$).  The Strehl ratios implied by these fits are 
10$\pm0.5$\% and 19$\pm1.0$\% , for $H$ and $K_{S}$ band, 
respectively.}
\label{pspec}
\end{figure}

The MTF-fitting technique has so far only been applied to data
obtained for the Keck Observatory LGSAO system \citep{Sheehy06,
Vacca07}.  To apply this technique to the VLT images, we used the
\citet{Sheehy06} software with appropriate input parameters for the
VLT telescope pupil size and geometry, the camera platescale and pixel
size, and the deformable mirror actuator spacing.  The software uses
the IDL procedure MPFIT\footnote{Available at
http://cow.physics.wisc.edu/~craigm/idl/.} to perform a
Levenberg-Marquardt least-squares fit of the parameterized model MTF
to the data.  The best fits to the $H$ and $K_{S}$ band image power
spectra are shown in Figure \ref{pspec}.  The spatial frequency,
$\nu_{n}$, is normalized relative to the telescope cutoff frequency in
the image plane (D$_{tel}/\lambda$).  

All images, even seeing-limited images, contain power up to the
spatial frequency cutoff, which is $D/\lambda$ for a circular pupil.
To faithfully record all spatial information allowed by the telescope
aperture, the pixel sampling frequency must therefore satisfy
$1/\delta x \ge 2D/\lambda$ (see \citet{Sheehy06} for a detailed
discussion).  The $K_{S}$ band data are Nyquist-sampled: $(1/\delta x)
/(2D/\lambda)$ = 1.  However, the H-band data are undersampled:
$(1/\delta x) /(2D/\lambda) = 0.76$.  This aliasing causes power at
frequencies $0.76 < \nu_{n} < 1$ to corrupt frequencies $0.52 <
\nu_{n} < 0.76$.  The contamination of frequencies between 0.52 and
$\sim$0.7 is negligible, as the power aliased to these frequencies
consists only of detector and background noise.  This power is orders
of magnitude less than the non-aliased power at these spatial
frequencies, and thus, has very little effect on the total power
spectrum.  On the other hand, the aliased power for contaminated
spatial frequencies $\nu_{n} \gtrsim 0.7$ is of the same
order-of-magnitude as the non-aliased power, so the power spectrum at
these frequencies is heavily influenced by the aliasing.  We,
therefore, chose only to fit the power spectrum for all spatial
frequencies $\nu_{n} < 0.70$. As only detector and background noise
contribute to the power spectrum for spatial frequencies $\nu_{n}
\gtrsim 0.8 $, we are excluding very little information about the PSF.

The relatively minor discrepency in Figure \ref{pspec} between the
model and data power spectra in $K_{S}$ band is likely due to
uncertainties in the deformable mirror influence function, as an
explicit description of the VLT CILAS mirror influence function was
not available.  We, therefore, used the approximate influence function
for the Keck Xinetics mirror from \citet{vanDam04}.  The overall power
spectra fits are satisfactory, and we do not expect that the
photometric accuracy of our technique is significantly affected by
this choice.

The Strehl ratios implied by the PSFs reconstructed from the fits in
Figure \ref{pspec} are 10\% and 19\%, for $H$ and $K_{S}$ band,
respectively.  To test the integrity of these fits, we divided the
images into nine subimages and fit the power spectrum of each
subimage.  This process yielded fairly consistent results for all
subimages, implying an error in the $H$ and $K_{S}$ band Strehl ratios
of 0.5\% and 1.0\%, respectively.  This agreement between subimages
suggests that PSF variations due to variable distance from the guide
star and variable exposure time in the final mosaic are only present
at a low level.

We determined the photometric zero point from observations of the
near-IR standard star 9178 \citep{Persson98}, obtained on the same
night.  The measured atmospheric extinction coefficients used to
correct for the difference in airmass between our standard stars and
science targets were 0.06 mag. airmass$^{-1}$ for $H$ and 0.07 mag. 
airmass$^{-1}$ for $K_{S}$.

\begin{figure}
\epsscale{0.7}
\includegraphics[width=.35\textwidth,angle=90]{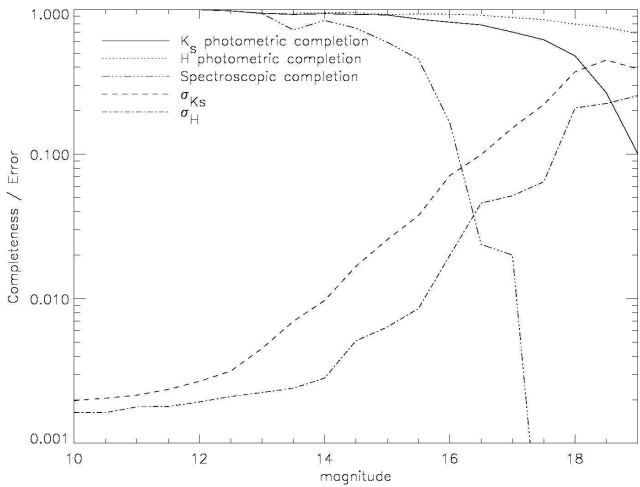}
\caption{$H$ and $K_{S}$ band photometric errors and completeness
based on the artificial star tests described in \S 2.  The
spectroscopic completeness is estimated based on the number of
CO-stars for which we could extract spectra, compared to the total
number of stars detected in the photometry.}
\label{comp_err}
\end{figure}

We derived photometric errors and completeness using the standard
technique of inserting and recovering artificial stars of specified
magnitude.  Completeness is not uniform, so we calculated errors and
completeness separately for each spectral field.  To avoid
artificially crowding our images, we inserted only fifty stars at a
time to each spectral subimage and repeated this procedure five times
for each magnitude bin.  We computed the photometric error at each
magnitude using a Gaussian fit to the difference between the input and
recovered magnitudes.  Figure \ref{comp_err} shows the average
photometric errors and completeness at each magnitude bin. Also shown
is the spectroscopic completeness, estimated by comparing the number
of stars detected in the $H$ and $K_S$ band photometry to the number
detected in the SINFONI data.  The incompleteness in the SINFONI data
at $K_{S}\sim$13-15 mag.  is caused by crowding and confusion with
nearby bright stars in the field of view.

\section{Results}

\subsection{Hertzsprung-Russell  Diagram Construction}

Using the derived photometry and effective temperatures presented in
the preceding section, we are able to place our GC sample on the
Hertzsprung-Russell (H-R) diagram.  To convert the measured
photometric magnitudes to luminosity, we use models that assume the
$K$ band magnitudes are measured in the Johnson-Cousins-Glass system
\citep{Bessell88}.  Therefore, we first transformed the measured
$K_{S}$ magnitudes to the Johnson-Cousins-Glass system, using the
approximate transformations given in \citet{Carpenter01}, assuming the
NaCo and 2MASS $JHK_{S}$ color systems are identical, as in
\citet{McCaughrean04}.  We next adopted empirical bolometric
corrections of \citet{Fluks94} for M stars and theoretical corrections
of Girardi 2005 (http://pleiadi/pd.astro.it) for hotter stars.  For
the theoretical corrections, we assumed a solar metallicity and a
surface gravity of $\log g = 2.0$, though the results are largely
independent of these assumptions.  For all temperatures in the
Galactic Center sample, we found systematic differences $< 0.05$
mag. in BC$_K$ for $\Delta[M/H]=1.0$ dex and $\Delta \log g=1.0$ dex,
leading to systematic differences in the luminosity of $\Delta \log(L/
L_{\sun}) \lesssim 0.02$. Additionally, we assumed a distance to the
Galactic Center of 8.0 kpc \citep{Reid93, Eisenhauer03c}.  There has
been much recent debate concerning the precise distance to the
Galactic Center.  However, the choice of distance has only a small
systematic effect on our derived luminosities; a change in distance of
$\Delta d = 0.4$ kpc induces a change in luminosity of $\Delta \log(L/
L_{\sun}) \sim 0.05$.  Finally, we corrected each star's luminosity
individually for extinction using the photometric color, the derived
\teff, and the interstellar extinction law of \citet{Rieke99}, derived
from NICMOS observations of Galactic Center stars.  All results are
tabulated in Table \ref{gc_table}.

\begin{longtable*}{lccccccccccccc}
\tablecaption{Properties of the Late-type Galactic Center Stars}
\tabletypesize{\scriptsize}
\tablecolumns{14}
\tablewidth{0pt}
\tablehead{
\colhead{ID} & \colhead{$\Delta$RA} & \colhead{$\Delta$Dec} & \colhead{$K$} & \colhead{$\sigma_K$} & \colhead{$H$} & \colhead{$\sigma_H$} &
\colhead{CO} & \colhead{$\log$ \teff} & \colhead{$\sigma_{\log T_{eff}}$} & \colhead{$A_{K}$} & \colhead{$\sigma_{A_{K}}$} &
\colhead{$\log(L/ L_{\sun})$} & \colhead{$\sigma_{\log(L/ L_{\sun})}$} \\
 & ($''$) & ($''$) & (mag) & (mag) & (mag) & (mag) &
(\%) & & & (mag) & (mag) & &
}
       1 & 6.26 & 6.24 & 15.275 & 0.042 & 17.419 & 0.072 & 8.28 & 3.658 & 0.006
 & 2.99 & 0.15 & 1.88 & 0.06 \\ 
       2 & 6.32 & 6.05 & 15.259 & 0.042 & 17.316 & 0.072 & 8.30 & 3.658 & 0.007
 & 2.86 & 0.15 & 1.83 & 0.07 \\ 
       3 & 5.67 & 5.91 & 14.898 & 0.031 & 17.051 & 0.056 & 6.93 & 3.674 & 0.004
 & 3.03 & 0.11 & 2.09 & 0.05 \\ 
       4 & 5.67 & 6.13 & 14.239 & 0.013 & 16.344 & 0.049 & 8.61 & 3.654 & 0.006
 & 2.93 & 0.09 & 2.25 & 0.04 \\ 
       5 & 5.51 & 6.24 & 15.304 & 0.042 & 17.438 & 0.072 & 7.01 & 3.673 & 0.005
 & 3.00 & 0.15 & 1.91 & 0.06 \\ 
       6 & 5.02 & 6.48 & 15.531 & 0.042 & 17.826 & 0.204 & 7.15 & 3.672 & 0.008
 & 3.23 & 0.37 & 1.91 & 0.15 \\ 
       7 & 4.86 & 6.29 & 15.131 & 0.031 & 17.274 & 0.072 & 7.01 & 3.673 & 0.007
 & 3.01 & 0.14 & 1.98 & 0.06 \\ 
       8 & 4.99 & 6.16 & 15.327 & 0.042 & 17.406 & 0.072 & 7.34 & 3.669 & 0.005
 & 2.91 & 0.15 & 1.86 & 0.06 \\ 
       9 & 4.75 & 6.13 & 15.503 & 0.042 & 17.678 & 0.072 & 8.76 & 3.652 & 0.006
 & 3.03 & 0.15 & 1.78 & 0.06 \\ 
      10 & 4.35 & 6.34 & 15.474 & 0.042 & 17.135 & 0.056 & 6.82 & 3.675 & 0.004
 & 2.30 & 0.12 & 1.57 & 0.05 \\ 
      11 & 4.35 & 6.51 & 15.240 & 0.031 & 17.387 & 0.072 & 6.69 & 3.677 & 0.004
 & 3.02 & 0.14 & 1.96 & 0.06 \\ 
      12 & 3.54 & 6.56 & 14.318 & 0.017 & 16.427 & 0.049 & 7.73 & 3.665 & 0.005
 & 2.95 & 0.09 & 2.26 & 0.04 \\ 
      13 & 3.56 & 6.40 & 13.803 & 0.013 & 15.707 & 0.012 & 9.40 & 3.643 & 0.005
 & 2.62 & 0.03 & 2.27 & 0.02 \\ 
      14 & 3.48 & 6.13 & 14.969 & 0.031 & 16.899 & 0.056 & 6.90 & 3.675 & 0.005
 & 2.70 & 0.11 & 1.93 & 0.05 \\ 
      15 & 3.08 & 6.13 & 12.132 & 0.003 & 14.305 & 0.006 & 11.62 & 3.611 & 0.005
 & 2.96 & 0.02 & 2.99 & 0.01 \\ 
      16 & 3.05 & 5.83 & 14.123 & 0.013 & 16.226 & 0.042 & 8.65 & 3.653 & 0.006
 & 2.93 & 0.08 & 2.30 & 0.04 \\ 
      17 & 2.81 & 5.86 & 15.867 & 0.078 & 17.802 & 0.204 & 8.77 & 3.652 & 0.005
 & 2.67 & 0.39 & 1.49 & 0.16 \\ 
      18 & 3.27 & 5.62 & 15.596 & 0.042 & 17.751 & 0.204 & 9.25 & 3.645 & 0.007
 & 2.99 & 0.37 & 1.71 & 0.15 \\ 
      19 & 2.78 & 5.32 & 15.115 & 0.031 & 17.234 & 0.056 & 9.20 & 3.646 & 0.007
 & 2.94 & 0.11 & 1.88 & 0.05 \\ 
      20 & 4.35 & 5.18 & 11.399 & 0.002 & 13.519 & 0.003 & 12.10 & 3.604 & 0.005
 & 2.87 & 0.01 & 3.23 & 0.01 \\ 
      21 & 4.78 & 5.37 & 14.438 & 0.017 & 16.735 & 0.049 & 9.71 & 3.639 & 0.004
 & 3.19 & 0.09 & 2.24 & 0.04 \\ 
      22 & 5.13 & 5.51 & 15.436 & 0.042 & 17.641 & 0.072 & 8.55 & 3.655 & 0.004
 & 3.08 & 0.15 & 1.84 & 0.06 \\ 
      23 & 4.97 & 5.10 & 15.341 & 0.042 & 17.335 & 0.072 & 10.09 & 3.634 & 0.005
 & 2.74 & 0.15 & 1.68 & 0.06 \\ 
      24 & 5.53 & 4.54 & 14.770 & 0.031 & 17.054 & 0.056 & 6.77 & 3.676 & 0.006
 & 3.22 & 0.11 & 2.22 & 0.05 \\ 
      25 & 5.10 & 3.46 & 14.078 & 0.013 & 15.666 & 0.012 & 8.31 & 3.658 & 0.005
 & 2.17 & 0.03 & 2.02 & 0.02 \\ 
      26 & 5.16 & 3.67 & 15.087 & 0.031 & 17.025 & 0.056 & 9.92 & 3.636 & 0.011
 & 2.66 & 0.12 & 1.76 & 0.06 \\ 
      27 & 5.45 & 3.91 & 15.316 & 0.042 & 17.369 & 0.072 & 10.07 & 3.634 & 0.007
 & 2.82 & 0.15 & 1.73 & 0.06 \\ 
      28 & 5.08 & 4.18 & 13.152 & 0.005 & 15.121 & 0.006 & 10.32 & 3.631 & 0.006
 & 2.69 & 0.02 & 2.53 & 0.02 \\ 
      29 & 5.13 & 4.51 & 13.238 & 0.005 & 15.345 & 0.012 & 10.91 & 3.622 & 0.005
 & 2.88 & 0.02 & 2.55 & 0.02 \\ 
      30 & 5.05 & 4.70 & 13.980 & 0.013 & 16.298 & 0.049 & 12.37 & 3.600 & 0.007
 & 3.16 & 0.09 & 2.30 & 0.04 \\ 
      31 & 4.56 & 4.56 & 13.382 & 0.009 & 15.542 & 0.012 & 8.63 & 3.653 & 0.004
 & 3.01 & 0.03 & 2.63 & 0.02 \\ 
      32 & 4.29 & 4.29 & 15.536 & 0.042 & 17.606 & 0.072 & 7.99 & 3.662 & 0.004
 & 2.89 & 0.15 & 1.74 & 0.06 \\ 
      33 & 4.59 & 4.13 & 16.137 & 0.078 & 18.223 & 0.204 & 5.27 & 3.692 & 0.003
 & 2.95 & 0.39 & 1.61 & 0.16 \\ 
      34 & 4.40 & 3.48 & 13.845 & 0.013 & 15.739 & 0.012 & 8.65 & 3.653 & 0.006
 & 2.62 & 0.03 & 2.28 & 0.02 \\ 
      35 & 4.24 & 3.70 & 13.676 & 0.009 & 15.619 & 0.012 & 7.61 & 3.666 & 0.005
 & 2.71 & 0.03 & 2.42 & 0.02 \\ 
      36 & 4.32 & 3.02 & 13.194 & 0.005 & 15.295 & 0.012 & 9.46 & 3.642 & 0.005
 & 2.91 & 0.02 & 2.63 & 0.02 \\ 
      37 & 3.00 & 3.32 & 12.562 & 0.003 & 14.834 & 0.006 & 3.70 & 3.707 & 0.004
 & 3.24 & 0.01 & 3.20 & 0.01 \\ 
      38 & 3.11 & 3.83 & 15.976 & 0.078 & 18.227 & 0.204 & 9.31 & 3.645 & 0.008
 & 3.13 & 0.39 & 1.61 & 0.16 \\ 
      39 & 3.62 & 3.56 & 15.106 & 0.031 & 17.234 & 0.056 & 5.89 & 3.686 & 0.004
 & 3.00 & 0.11 & 2.03 & 0.05 \\ 
      40 & 4.02 & 3.97 & 15.532 & 0.042 & 17.603 & 0.072 & 7.97 & 3.662 & 0.006
 & 2.89 & 0.15 & 1.74 & 0.06 \\ 
      41 & 3.62 & 4.05 & 15.160 & 0.031 & 17.431 & 0.072 & 9.04 & 3.648 & 0.005
 & 3.17 & 0.14 & 1.96 & 0.06 \\ 
      42 & 2.92 & 4.13 & 12.738 & 0.003 & 15.117 & 0.006 & 12.73 & 3.594 & 0.005
 & 3.23 & 0.02 & 2.82 & 0.02 \\ 
      43 & -6.97 & 10.15 & 15.030 & 0.026 & 16.988 & 0.058 & 8.98 & 3.649 & 
0.008 & 2.71 & 0.11 & 1.83 & 0.05 \\ 
      44 & -7.24 & 10.31 & 14.751 & 0.026 & 16.765 & 0.058 & 7.28 & 3.670 & 
0.006 & 2.82 & 0.11 & 2.05 & 0.05 \\ 
      45 & -7.10 & 9.85 & 15.112 & 0.026 & 17.048 & 0.058 & 5.96 & 3.685 & 0.006
 & 2.72 & 0.11 & 1.91 & 0.05 \\ 
      46 & -7.32 & 9.88 & 15.075 & 0.026 & 17.022 & 0.058 & 6.46 & 3.680 & 0.006
 & 2.73 & 0.11 & 1.91 & 0.05 \\ 
      47 & -7.94 & 9.88 & 15.494 & 0.040 & 17.295 & 0.060 & 4.70 & 3.698 & 0.006
 & 2.53 & 0.13 & 1.72 & 0.06 \\ 
      48 & -8.99 & 10.12 & 13.124 & 0.005 & 15.144 & 0.008 & 10.30 & 3.631 & 
0.008 & 2.77 & 0.02 & 2.57 & 0.02 \\ 
      49 & -8.80 & 9.67 & 15.374 & 0.040 & 17.408 & 0.060 & 7.61 & 3.666 & 0.009
 & 2.84 & 0.13 & 1.80 & 0.06 \\ 
      50 & -9.29 & 9.37 & 15.199 & 0.026 & 17.210 & 0.058 & 8.72 & 3.652 & 0.007
 & 2.79 & 0.11 & 1.81 & 0.05 \\ 
      51 & -8.69 & 9.21 & 15.156 & 0.026 & 17.093 & 0.058 & 6.25 & 3.682 & 0.006
 & 2.71 & 0.11 & 1.88 & 0.05 \\ 
      52 & -7.45 & 9.37 & 11.937 & 0.003 & 13.686 & 0.002 & 5.88 & 3.686 & 0.004
 & 2.44 & 0.01 & 3.07 & 0.01 \\ 
      53 & -5.59 & 9.21 & 10.319 & 0.002 & 12.222 & 0.002 & 11.08 & 3.619 & 
0.008 & 2.58 & 0.02 & 3.59 & 0.02 \\ 
      54 & -6.75 & 8.80 & 15.448 & 0.040 & 17.392 & 0.060 & 6.35 & 3.681 & 0.005
 & 2.72 & 0.13 & 1.76 & 0.06 \\ 
      55 & -6.75 & 8.56 & 15.948 & 0.090 & 18.090 & 0.245 & 6.47 & 3.679 & 0.008
 & 3.02 & 0.46 & 1.68 & 0.19 \\ 
      56 & -6.18 & 8.18 & 15.273 & 0.040 & 17.416 & 0.060 & 8.85 & 3.651 & 0.009
 & 2.98 & 0.13 & 1.85 & 0.06 \\ 
      57 & -6.29 & 8.07 & 15.706 & 0.040 & 17.964 & 0.245 & 8.91 & 3.650 & 0.007
 & 3.15 & 0.44 & 1.74 & 0.18 \\ 
      58 & -6.10 & 7.86 & 15.506 & 0.040 & 17.688 & 0.060 & 4.64 & 3.699 & 0.005
 & 3.10 & 0.13 & 1.94 & 0.06 \\ 
      59 & -5.97 & 7.29 & 15.037 & 0.026 & 17.188 & 0.058 & 4.32 & 3.702 & 0.008
 & 3.05 & 0.11 & 2.12 & 0.05 \\ 
      60 & -6.16 & 6.43 & 13.182 & 0.005 & 15.300 & 0.009 & 7.56 & 3.667 & 0.008
 & 2.97 & 0.02 & 2.73 & 0.02 \\ 
      61 & -6.51 & 6.64 & 15.628 & 0.040 & 17.772 & 0.245 & 9.21 & 3.646 & 0.008
 & 2.98 & 0.44 & 1.69 & 0.18 \\ 
      62 & -6.80 & 6.86 & 14.604 & 0.022 & 16.766 & 0.058 & 7.46 & 3.668 & 0.006
 & 3.03 & 0.11 & 2.19 & 0.05 \\ 
      63 & -7.32 & 7.78 & 13.190 & 0.005 & 15.195 & 0.008 & 8.64 & 3.653 & 0.006
 & 2.78 & 0.02 & 2.61 & 0.02 \\ 
      64 & -7.45 & 8.29 & 15.413 & 0.040 & 17.374 & 0.060 & 6.44 & 3.680 & 0.005
 & 2.75 & 0.13 & 1.79 & 0.06 \\ 
      65 & -7.64 & 8.56 & 15.467 & 0.040 & 17.477 & 0.060 & 6.86 & 3.675 & 0.006
 & 2.82 & 0.13 & 1.78 & 0.06 \\ 
      66 & -7.48 & 8.99 & 12.976 & 0.005 & 14.980 & 0.008 & 12.14 & 3.603 & 
0.009 & 2.70 & 0.03 & 2.53 & 0.02 \\ 
      67 & -8.23 & 8.88 & 15.498 & 0.040 & 17.511 & 0.060 & 5.75 & 3.687 & 0.008
 & 2.83 & 0.13 & 1.81 & 0.06 \\ 
      68 & -8.91 & 8.64 & 15.444 & 0.040 & 17.651 & 0.060 & 9.59 & 3.641 & 0.008
 & 3.06 & 0.13 & 1.79 & 0.06 \\ 
      69 & -8.59 & 8.18 & 12.292 & 0.004 & 14.300 & 0.005 & 8.80 & 3.651 & 0.005
 & 2.78 & 0.01 & 2.97 & 0.02 \\ 
      70 & -8.13 & 8.13 & 14.233 & 0.010 & 16.282 & 0.047 & 8.98 & 3.649 & 0.009
 & 2.84 & 0.09 & 2.21 & 0.04 \\ 
      71 & -8.18 & 8.32 & 15.497 & 0.040 & 17.501 & 0.060 & 7.83 & 3.664 & 0.007
 & 2.79 & 0.13 & 1.72 & 0.06 \\ 
      72 & -7.94 & 8.45 & 15.375 & 0.040 & 17.420 & 0.060 & 9.34 & 3.644 & 0.007
 & 2.83 & 0.13 & 1.73 & 0.06 \\ 
      73 & -7.75 & 8.26 & 15.616 & 0.040 & 17.814 & 0.245 & 6.29 & 3.682 & 0.006
 & 3.10 & 0.44 & 1.85 & 0.18 \\ 
      74 & -7.97 & 7.51 & 12.569 & 0.004 & 14.867 & 0.008 & 10.02 & 3.635 & 
0.009 & 3.19 & 0.02 & 2.97 & 0.02 \\ 
      75 & -7.61 & 7.21 & 14.774 & 0.026 & 16.954 & 0.058 & 5.67 & 3.688 & 0.007
 & 3.08 & 0.11 & 2.20 & 0.05 \\ 
      76 & -8.45 & 6.99 & 10.843 & 0.002 & 13.278 & 0.002 & 12.81 & 3.593 & 
0.013 & 3.32 & 0.03 & 3.60 & 0.03 \\ 
      77 & -9.21 & 6.53 & 12.543 & 0.004 & 14.852 & 0.008 & 10.86 & 3.623 & 
0.008 & 3.18 & 0.02 & 2.95 & 0.02 \\ 
      78 & -8.64 & 7.34 & 15.402 & 0.040 & 17.464 & 0.060 & 8.56 & 3.654 & 0.007
 & 2.87 & 0.13 & 1.76 & 0.06 \\ 
      79 & 2.94 & 12.93 & 15.314 & 0.049 & 16.938 & 0.055 & 5.66 & 3.688 & 0.004
 & 2.26 & 0.13 & 1.65 & 0.06 \\ 
      80 & 2.67 & 12.74 & 14.953 & 0.027 & 16.828 & 0.055 & 7.95 & 3.662 & 0.005
 & 2.60 & 0.11 & 1.86 & 0.05 \\ 
      81 & 2.54 & 12.45 & 15.140 & 0.027 & 16.947 & 0.055 & 8.00 & 3.661 & 0.006
 & 2.50 & 0.11 & 1.74 & 0.05 \\ 
      82 & 2.59 & 12.04 & 15.189 & 0.027 & 16.632 & 0.046 & 4.90 & 3.696 & 0.005
 & 2.00 & 0.09 & 1.62 & 0.04 \\ 
      83 & 3.21 & 12.69 & 16.003 & 0.075 & 17.874 & 0.210 & 7.56 & 3.667 & 0.007
 & 2.60 & 0.40 & 1.45 & 0.16 \\ 
      84 & 3.00 & 12.64 & 15.868 & 0.075 & 17.752 & 0.210 & 6.33 & 3.681 & 0.006
 & 2.64 & 0.40 & 1.56 & 0.16 \\ 
      85 & 5.62 & 11.88 & 15.501 & 0.049 & 17.546 & 0.080 & 9.55 & 3.641 & 0.006
 & 2.82 & 0.17 & 1.67 & 0.07 \\ 
      86 & 5.83 & 11.42 & 11.834 & 0.003 & 14.160 & 0.003 & 13.30 & 3.585 & 
0.007 & 3.14 & 0.02 & 3.11 & 0.02 \\ 
      87 & 5.94 & 11.12 & 13.811 & 0.011 & 16.328 & 0.046 & 10.99 & 3.621 & 
0.008 & 3.49 & 0.09 & 2.56 & 0.04 \\ 
      88 & 5.48 & 11.45 & 13.479 & 0.008 & 16.289 & 0.046 & 11.75 & 3.609 & 
0.007 & 3.90 & 0.08 & 2.83 & 0.04 \\ 
      89 & 5.18 & 11.53 & 15.624 & 0.049 & 17.880 & 0.210 & 6.90 & 3.675 & 0.005
 & 3.18 & 0.38 & 1.86 & 0.16 \\ 
      90 & 4.83 & 11.74 & 15.531 & 0.049 & 17.641 & 0.080 & 7.38 & 3.669 & 0.005
 & 2.96 & 0.17 & 1.79 & 0.07 \\ 
      91 & 4.78 & 11.53 & 15.510 & 0.049 & 17.643 & 0.080 & 4.91 & 3.696 & 0.006
 & 3.02 & 0.17 & 1.90 & 0.07 \\ 
      92 & 4.35 & 11.69 & 14.863 & 0.027 & 16.784 & 0.055 & 8.03 & 3.661 & 0.005
 & 2.67 & 0.11 & 1.92 & 0.05 \\ 
      93 & 3.70 & 11.42 & 12.593 & 0.003 & 14.495 & 0.006 & 5.95 & 3.685 & 0.003
 & 2.67 & 0.01 & 2.90 & 0.01 \\ 
      94 & 3.43 & 11.39 & 13.861 & 0.011 & 15.757 & 0.042 & 8.13 & 3.660 & 0.005
 & 2.63 & 0.08 & 2.30 & 0.03 \\ 
      95 & 2.38 & 10.96 & 12.391 & 0.003 & 14.350 & 0.006 & 11.31 & 3.616 & 
0.004 & 2.66 & 0.01 & 2.78 & 0.01 \\ 
      96 & 2.59 & 10.77 & 13.164 & 0.005 & 14.799 & 0.006 & 6.53 & 3.679 & 0.004
 & 2.26 & 0.01 & 2.49 & 0.01 \\ 
      97 & 3.16 & 10.69 & 14.694 & 0.019 & 16.701 & 0.046 & 5.54 & 3.690 & 0.006
 & 2.83 & 0.09 & 2.13 & 0.04 \\ 
      98 & 3.73 & 10.53 & 12.842 & 0.005 & 14.355 & 0.006 & 8.50 & 3.655 & 0.005
 & 2.06 & 0.01 & 2.47 & 0.01 \\ 
      99 & 4.51 & 10.85 & 11.905 & 0.003 & 13.794 & 0.003 & 5.57 & 3.689 & 0.003
 & 2.65 & 0.01 & 3.18 & 0.01 \\ 
     100 & 5.10 & 10.66 & 13.593 & 0.008 & 15.708 & 0.008 & 9.17 & 3.646 & 0.007
 & 2.93 & 0.02 & 2.49 & 0.02 \\ 
     101 & 5.67 & 10.80 & 15.151 & 0.027 & 17.216 & 0.055 & 8.89 & 3.650 & 0.005
 & 2.86 & 0.11 & 1.85 & 0.05 \\ 
     102 & 5.40 & 9.96 & 15.114 & 0.027 & 17.121 & 0.055 & 6.77 & 3.676 & 0.006
 & 2.81 & 0.11 & 1.92 & 0.05 \\ 
     103 & 5.16 & 9.88 & 15.134 & 0.027 & 17.133 & 0.055 & 8.75 & 3.652 & 0.007
 & 2.77 & 0.11 & 1.83 & 0.05 \\ 
     104 & 5.91 & 9.45 & 15.730 & 0.049 & 17.717 & 0.080 & 6.20 & 3.682 & 0.009
 & 2.79 & 0.17 & 1.68 & 0.07 \\ 
     105 & 5.08 & 10.18 & 13.307 & 0.008 & 15.413 & 0.008 & 11.02 & 3.620 & 
0.007 & 2.88 & 0.02 & 2.51 & 0.02 \\ 
     106 & 3.59 & 9.45 & 13.281 & 0.008 & 15.303 & 0.008 & 11.49 & 3.613 & 0.006
 & 2.74 & 0.02 & 2.45 & 0.02 \\ 
     107 & 3.29 & 9.40 & 15.212 & 0.027 & 17.153 & 0.055 & 7.64 & 3.666 & 0.009
 & 2.70 & 0.11 & 1.81 & 0.05 \\ 
     108 & 3.13 & 9.53 & 15.129 & 0.027 & 16.961 & 0.055 & 6.68 & 3.677 & 0.005
 & 2.55 & 0.11 & 1.81 & 0.05 \\ 
     109 & 3.08 & 9.83 & 13.819 & 0.011 & 15.715 & 0.008 & 9.31 & 3.644 & 0.006
 & 2.61 & 0.03 & 2.27 & 0.02 \\ 
     110 & 3.46 & 10.12 & 13.302 & 0.008 & 15.255 & 0.008 & 14.05 & 3.572 & 
0.007 & 2.56 & 0.03 & 2.27 & 0.02 \\ 
     111 & 2.94 & 10.07 & 14.905 & 0.027 & 16.849 & 0.055 & 8.69 & 3.653 & 0.005
 & 2.69 & 0.11 & 1.89 & 0.05 \\ 
     112 & 2.51 & 10.23 & 14.381 & 0.019 & 15.823 & 0.042 & 5.62 & 3.689 & 0.005
 & 1.99 & 0.08 & 1.92 & 0.04 \\ 
     113 & 2.67 & 9.53 & 15.518 & 0.049 & 17.613 & 0.080 & 7.98 & 3.662 & 0.007
 & 2.92 & 0.17 & 1.76 & 0.07 \\ 
     114 & -4.86 & 12.72 & 15.647 & 0.042 & 17.899 & 0.224 & 7.68 & 3.665 & 
0.007 & 3.16 & 0.41 & 1.82 & 0.16 \\ 
     115 & -5.40 & 12.39 & 14.498 & 0.019 & 16.366 & 0.048 & 7.14 & 3.672 & 
0.005 & 2.60 & 0.09 & 2.07 & 0.04 \\ 
     116 & -6.18 & 12.69 & 15.679 & 0.042 & 17.406 & 0.072 & 6.76 & 3.676 & 
0.005 & 2.40 & 0.15 & 1.53 & 0.06 \\ 
     117 & -6.59 & 12.77 & 16.205 & 0.068 & 18.072 & 0.224 & 5.31 & 3.692 & 
0.005 & 2.62 & 0.42 & 1.45 & 0.17 \\ 
     118 & -6.78 & 12.66 & 16.404 & 0.119 & 18.180 & 0.224 & 6.30 & 3.681 & 
0.008 & 2.48 & 0.45 & 1.28 & 0.19 \\ 
     119 & -7.48 & 12.26 & 15.019 & 0.033 & 16.737 & 0.048 & 7.69 & 3.665 & 
0.005 & 2.37 & 0.10 & 1.75 & 0.05 \\ 
     120 & -7.70 & 12.45 & 15.224 & 0.033 & 16.988 & 0.053 & 6.94 & 3.674 & 
0.005 & 2.45 & 0.11 & 1.73 & 0.05 \\ 
     121 & -7.07 & 11.93 & 15.166 & 0.033 & 16.674 & 0.048 & 5.40 & 3.691 & 
0.004 & 2.09 & 0.10 & 1.65 & 0.04 \\ 
     122 & -6.86 & 11.85 & 15.556 & 0.042 & 17.281 & 0.072 & 6.11 & 3.683 & 
0.004 & 2.40 & 0.15 & 1.60 & 0.06 \\ 
     123 & -6.56 & 11.99 & 15.669 & 0.042 & 17.310 & 0.072 & 5.47 & 3.690 & 
0.004 & 2.29 & 0.15 & 1.53 & 0.06 \\ 
     124 & -6.21 & 12.10 & 15.638 & 0.042 & 17.358 & 0.072 & 6.14 & 3.683 & 
0.006 & 2.40 & 0.15 & 1.56 & 0.06 \\ 
     125 & -5.16 & 12.20 & 17.135 & 0.131 & 19.005 & 0.252 & 6.87 & 3.675 & 
0.005 & 2.61 & 0.51 & 1.03 & 0.21 \\ 
     126 & -4.89 & 12.12 & 15.325 & 0.042 & 17.257 & 0.072 & 5.98 & 3.685 & 
0.005 & 2.71 & 0.15 & 1.82 & 0.06 \\ 
     127 & -5.18 & 11.85 & 15.313 & 0.042 & 17.172 & 0.053 & 5.02 & 3.695 & 
0.004 & 2.62 & 0.12 & 1.82 & 0.05 \\ 
     128 & -5.83 & 11.56 & 14.756 & 0.033 & 16.508 & 0.048 & 7.64 & 3.666 & 
0.004 & 2.42 & 0.10 & 1.88 & 0.05 \\ 
     129 & -6.56 & 11.45 & 14.853 & 0.033 & 16.636 & 0.048 & 8.90 & 3.650 & 
0.005 & 2.45 & 0.10 & 1.80 & 0.05 \\ 
     130 & -7.21 & 11.53 & 13.517 & 0.008 & 15.305 & 0.010 & 9.99 & 3.635 & 
0.005 & 2.43 & 0.02 & 2.29 & 0.02 \\ 
     131 & -7.51 & 11.74 & 14.319 & 0.019 & 16.096 & 0.045 & 7.41 & 3.669 & 
0.004 & 2.46 & 0.09 & 2.08 & 0.04 \\ 
     132 & -7.75 & 11.58 & 15.081 & 0.033 & 16.952 & 0.053 & 8.21 & 3.659 & 
0.007 & 2.59 & 0.11 & 1.80 & 0.05 \\ 
     133 & -8.23 & 11.45 & 15.838 & 0.068 & 17.605 & 0.072 & 7.01 & 3.673 & 
0.004 & 2.46 & 0.18 & 1.48 & 0.08 \\ 
     134 & -8.29 & 11.12 & 15.889 & 0.068 & 17.677 & 0.072 & 6.84 & 3.675 & 
0.004 & 2.49 & 0.18 & 1.48 & 0.08 \\ 
     135 & -7.80 & 11.18 & 14.080 & 0.011 & 15.992 & 0.045 & 10.11 & 3.634 & 
0.005 & 2.62 & 0.08 & 2.14 & 0.04 \\ 
     136 & -7.70 & 10.83 & 15.377 & 0.042 & 17.235 & 0.053 & 6.32 & 3.681 & 
0.005 & 2.60 & 0.12 & 1.74 & 0.05 \\ 
     137 & -7.32 & 11.29 & 14.858 & 0.033 & 16.667 & 0.048 & 7.69 & 3.665 & 
0.004 & 2.51 & 0.10 & 1.87 & 0.05 \\ 
     138 & -7.07 & 10.85 & 15.836 & 0.068 & 17.186 & 0.053 & 5.97 & 3.685 & 
0.006 & 1.85 & 0.15 & 1.27 & 0.07 \\ 
     139 & -6.83 & 10.88 & 14.494 & 0.019 & 16.366 & 0.048 & 10.31 & 3.631 & 
0.006 & 2.55 & 0.09 & 1.94 & 0.04 \\ 
     140 & -7.02 & 10.53 & 14.002 & 0.011 & 16.085 & 0.045 & 9.61 & 3.641 & 
0.005 & 2.88 & 0.08 & 2.29 & 0.04 \\ 
     141 & -7.26 & 10.50 & 14.588 & 0.019 & 16.339 & 0.048 & 7.36 & 3.669 & 
0.005 & 2.43 & 0.09 & 1.96 & 0.04 \\ 
     142 & -5.99 & 11.18 & 11.861 & 0.003 & 13.400 & 0.002 & 10.47 & 3.628 & 
0.005 & 2.06 & 0.01 & 2.79 & 0.01 \\ 
     143 & -5.89 & 10.66 & 13.946 & 0.011 & 15.939 & 0.045 & 8.36 & 3.657 & 
0.005 & 2.77 & 0.08 & 2.31 & 0.04 \\ 
     144 & -5.45 & 10.31 & 15.461 & 0.042 & 17.386 & 0.072 & 5.11 & 3.694 & 
0.004 & 2.71 & 0.15 & 1.79 & 0.06 \\ 
     145 & -5.26 & 10.18 & 14.684 & 0.019 & 16.372 & 0.048 & 7.67 & 3.666 & 
0.004 & 2.33 & 0.09 & 1.87 & 0.04 \\ 
     146 & -4.86 & 9.85 & 12.872 & 0.004 & 14.806 & 0.007 & 10.37 & 3.630 & 
0.005 & 2.64 & 0.02 & 2.62 & 0.01 \\ 
     147 & 4.94 & 16.85 & 15.552 & 0.033 & 17.771 & 0.193 & 8.39 & 3.657 & 0.007
 & 3.10 & 0.35 & 1.80 & 0.14 \\ 
     148 & 4.18 & 16.82 & 14.313 & 0.014 & 16.510 & 0.043 & 9.01 & 3.649 & 0.006
 & 3.06 & 0.08 & 2.26 & 0.04 \\ 
     149 & 3.51 & 16.69 & 15.608 & 0.033 & 17.420 & 0.058 & 7.22 & 3.671 & 0.005
 & 2.52 & 0.12 & 1.59 & 0.05 \\ 
     150 & 5.94 & 16.58 & 15.878 & 0.073 & 18.001 & 0.193 & 5.31 & 3.692 & 0.005
 & 3.00 & 0.37 & 1.74 & 0.15 \\ 
     151 & 5.83 & 16.28 & 15.617 & 0.033 & 17.657 & 0.058 & 7.14 & 3.672 & 0.006
 & 2.86 & 0.12 & 1.72 & 0.05 \\ 
     152 & 5.59 & 16.09 & 15.837 & 0.073 & 17.933 & 0.193 & 6.11 & 3.683 & 0.005
 & 2.95 & 0.37 & 1.71 & 0.15 \\ 
     153 & 4.97 & 16.42 & 12.691 & 0.003 & 15.117 & 0.005 & 14.50 & 3.564 & 
0.006 & 3.24 & 0.02 & 2.76 & 0.02 \\ 
     154 & 4.32 & 16.28 & 16.030 & 0.073 & 18.088 & 0.193 & 6.94 & 3.674 & 0.008
 & 2.89 & 0.37 & 1.58 & 0.15 \\ 
     155 & 2.73 & 16.60 & 15.914 & 0.073 & 17.716 & 0.058 & 5.83 & 3.686 & 0.008
 & 2.52 & 0.17 & 1.51 & 0.08 \\ 
     156 & 2.84 & 16.25 & 13.465 & 0.007 & 15.255 & 0.007 & 8.70 & 3.653 & 0.006
 & 2.46 & 0.02 & 2.37 & 0.02 \\ 
     157 & 3.00 & 15.98 & 15.508 & 0.033 & 17.399 & 0.058 & 4.92 & 3.696 & 0.006
 & 2.66 & 0.12 & 1.76 & 0.05 \\ 
     158 & 3.35 & 15.90 & 15.638 & 0.033 & 17.629 & 0.058 & 7.04 & 3.673 & 0.005
 & 2.79 & 0.12 & 1.69 & 0.05 \\ 
     159 & 3.73 & 15.96 & 15.501 & 0.033 & 17.434 & 0.058 & 6.61 & 3.678 & 0.006
 & 2.71 & 0.12 & 1.73 & 0.05 \\ 
     160 & 4.67 & 15.93 & 15.194 & 0.026 & 17.162 & 0.046 & 6.63 & 3.678 & 0.004
 & 2.76 & 0.09 & 1.87 & 0.04 \\ 
     161 & 5.32 & 15.77 & 15.439 & 0.033 & 17.512 & 0.058 & 7.90 & 3.663 & 0.005
 & 2.89 & 0.12 & 1.78 & 0.05 \\ 
     162 & 5.43 & 15.88 & 15.235 & 0.026 & 17.402 & 0.058 & 9.64 & 3.640 & 0.005
 & 3.00 & 0.11 & 1.85 & 0.05 \\ 
     163 & 5.78 & 15.80 & 15.236 & 0.026 & 17.399 & 0.058 & 8.57 & 3.654 & 0.007
 & 3.01 & 0.11 & 1.89 & 0.05 \\ 
     164 & 5.83 & 15.52 & 15.698 & 0.033 & 18.004 & 0.193 & 8.79 & 3.651 & 0.006
 & 3.22 & 0.35 & 1.78 & 0.14 \\ 
     165 & 5.99 & 15.58 & 16.165 & 0.073 & 18.239 & 0.193 & 6.62 & 3.678 & 0.008
 & 2.91 & 0.37 & 1.54 & 0.15 \\ 
     166 & 5.08 & 15.50 & 16.363 & 0.076 & 18.561 & 0.216 & 8.51 & 3.655 & 0.006
 & 3.07 & 0.41 & 1.46 & 0.17 \\ 
     167 & 4.99 & 15.15 & 15.308 & 0.033 & 17.408 & 0.058 & 10.85 & 3.623 & 
0.007 & 2.88 & 0.12 & 1.72 & 0.05 \\ 
     168 & 3.81 & 15.31 & 12.615 & 0.003 & 14.638 & 0.005 & 13.49 & 3.581 & 
0.007 & 2.68 & 0.02 & 2.61 & 0.02 \\ 
     169 & 3.38 & 15.31 & 15.383 & 0.033 & 17.319 & 0.058 & 8.35 & 3.657 & 0.006
 & 2.68 & 0.12 & 1.71 & 0.05 \\ 
     170 & 3.16 & 15.15 & 14.040 & 0.008 & 16.061 & 0.043 & 9.34 & 3.644 & 0.006
 & 2.79 & 0.08 & 2.25 & 0.04 \\ 
     171 & 2.92 & 15.66 & 15.052 & 0.026 & 16.976 & 0.046 & 8.92 & 3.650 & 0.008
 & 2.66 & 0.09 & 1.81 & 0.05 \\ 
     172 & 2.78 & 15.34 & 14.778 & 0.026 & 16.934 & 0.046 & 9.07 & 3.648 & 0.005
 & 3.00 & 0.09 & 2.05 & 0.04 \\ 
     173 & 2.54 & 14.80 & 13.166 & 0.004 & 14.934 & 0.005 & 5.00 & 3.695 & 0.005
 & 2.48 & 0.01 & 2.62 & 0.01 \\ 
     174 & 3.21 & 14.42 & 15.317 & 0.033 & 17.176 & 0.046 & 6.99 & 3.674 & 0.006
 & 2.59 & 0.10 & 1.74 & 0.05 \\ 
     175 & 3.24 & 14.66 & 15.590 & 0.033 & 17.545 & 0.058 & 7.46 & 3.668 & 0.006
 & 2.73 & 0.12 & 1.67 & 0.05 \\ 
     176 & 3.62 & 14.80 & 15.449 & 0.033 & 17.362 & 0.058 & 9.52 & 3.642 & 0.005
 & 2.63 & 0.12 & 1.61 & 0.05 \\ 
     177 & 4.10 & 14.88 & 15.905 & 0.073 & 18.090 & 0.193 & 7.66 & 3.666 & 0.005
 & 3.06 & 0.37 & 1.67 & 0.15 \\ 
     178 & 6.10 & 15.07 & 15.197 & 0.026 & 17.898 & 0.193 & 10.05 & 3.634 & 
0.006 & 3.78 & 0.35 & 2.16 & 0.14 \\ 
     179 & 5.75 & 14.50 & 10.803 & 0.002 & 13.327 & 0.002 & 16.04 & 3.536 & 
0.007 & 3.31 & 0.02 & 3.49 & 0.02 \\ 
     180 & 4.56 & 14.23 & 12.404 & 0.003 & 15.485 & 0.007 & 14.27 & 3.568 & 
0.007 & 4.22 & 0.02 & 3.28 & 0.02 \\ 
     181 & 4.00 & 14.28 & 15.637 & 0.033 & 17.484 & 0.058 & 4.80 & 3.697 & 0.003
 & 2.60 & 0.12 & 1.69 & 0.05 \\ 
     182 & 3.89 & 13.77 & 12.332 & 0.003 & 14.286 & 0.005 & 12.22 & 3.602 & 
0.007 & 2.62 & 0.02 & 2.75 & 0.02 \\ 
     183 & 3.67 & 14.09 & 15.154 & 0.026 & 17.022 & 0.046 & 9.60 & 3.641 & 0.005
 & 2.56 & 0.09 & 1.70 & 0.04 \\ 
     184 & 3.21 & 13.47 & 13.180 & 0.004 & 15.232 & 0.005 & 11.63 & 3.611 & 
0.007 & 2.78 & 0.02 & 2.50 & 0.02 \\ 
     185 & 3.00 & 14.09 & 15.071 & 0.026 & 17.010 & 0.046 & 7.33 & 3.670 & 0.005
 & 2.70 & 0.09 & 1.88 & 0.04 \\ 
     186 & 2.46 & 13.45 & 15.667 & 0.033 & 17.564 & 0.058 & 7.33 & 3.670 & 0.005
 & 2.64 & 0.12 & 1.61 & 0.05 \\ 
     187 & 4.75 & 13.82 & 15.838 & 0.073 & 17.896 & 0.193 & 7.32 & 3.670 & 0.004
 & 2.88 & 0.37 & 1.64 & 0.15 \\ 
     188 & 5.35 & 13.53 & 15.493 & 0.033 & 18.000 & 0.193 & 9.48 & 3.642 & 0.009
 & 3.51 & 0.35 & 1.95 & 0.14 \\ 
     189 & 5.21 & 13.01 & 13.400 & 0.007 & 15.575 & 0.007 & 11.59 & 3.612 & 
0.006 & 2.97 & 0.02 & 2.49 & 0.02 \\ 
     190 & 4.86 & 13.04 & 15.419 & 0.033 & 17.675 & 0.058 & 4.92 & 3.696 & 0.005
 & 3.20 & 0.12 & 2.01 & 0.05 \\ 
     191 & 4.62 & 13.28 & 15.571 & 0.033 & 17.613 & 0.058 & 7.31 & 3.670 & 0.005
 & 2.86 & 0.12 & 1.74 & 0.05 \\ 
     192 & 4.29 & 13.42 & 15.229 & 0.026 & 17.173 & 0.046 & 7.96 & 3.662 & 0.007
 & 2.70 & 0.09 & 1.79 & 0.04 \\ 
     193 & 4.13 & 13.18 & 15.177 & 0.026 & 17.135 & 0.046 & 5.54 & 3.690 & 0.006
 & 2.76 & 0.09 & 1.91 & 0.04 \\ 
     194 & 3.81 & 12.96 & 14.855 & 0.026 & 17.095 & 0.046 & 7.21 & 3.671 & 0.005
 & 3.15 & 0.09 & 2.14 & 0.04 \\ 
     195 & 3.75 & 12.82 & 14.958 & 0.026 & 16.816 & 0.046 & 6.62 & 3.678 & 0.006
 & 2.59 & 0.09 & 1.90 & 0.04 \\ 
     196 & 3.46 & 12.18 & 14.986 & 0.026 & 16.925 & 0.046 & 7.58 & 3.667 & 0.007
 & 2.70 & 0.09 & 1.90 & 0.04 \\ 
     197 & 4.62 & 12.31 & 14.028 & 0.008 & 16.714 & 0.043 & 11.82 & 3.608 & 
0.007 & 3.72 & 0.08 & 2.53 & 0.04 \\ 
     198 & 4.67 & 11.96 & 15.444 & 0.033 & 17.517 & 0.058 & 8.54 & 3.655 & 0.011
 & 2.88 & 0.12 & 1.76 & 0.06 \\ 
     199 & 5.48 & 12.34 & 14.066 & 0.008 & 16.048 & 0.043 & 9.55 & 3.641 & 0.009
 & 2.73 & 0.08 & 2.21 & 0.04 \\ 
     200 & 5.75 & 11.96 & 15.084 & 0.026 & 18.396 & 0.216 & 10.30 & 3.631 & 
0.010 & 4.68 & 0.39 & 2.55 & 0.16 \\ 
     201 & 5.83 & 12.50 & 15.007 & 0.026 & 16.824 & 0.046 & 7.83 & 3.664 & 0.007
 & 2.52 & 0.09 & 1.81 & 0.04 \\ 
     202 & 5.56 & 12.80 & 15.206 & 0.026 & 17.276 & 0.058 & 7.35 & 3.669 & 0.009
 & 2.90 & 0.11 & 1.90 & 0.05 \\ 
     203 & 5.32 & 12.64 & 15.004 & 0.026 & 16.793 & 0.046 & 7.93 & 3.662 & 0.006
 & 2.47 & 0.09 & 1.79 & 0.04 \\ 
     204 & 17.98 & 18.79 & 13.875 & 0.008 & 15.987 & 0.041 & 12.13 & 3.603 & 
0.006 & 2.86 & 0.08 & 2.23 & 0.03 \\ 
     205 & 17.42 & 18.66 & 12.761 & 0.005 & 14.750 & 0.005 & 7.67 & 3.665 & 
0.004 & 2.77 & 0.01 & 2.82 & 0.01 \\ 
     206 & 16.96 & 18.36 & 15.226 & 0.021 & 17.257 & 0.058 & 8.52 & 3.655 & 
0.005 & 2.82 & 0.11 & 1.82 & 0.05 \\ 
     207 & 16.63 & 18.68 & 14.970 & 0.021 & 16.876 & 0.048 & 8.24 & 3.658 & 
0.005 & 2.64 & 0.09 & 1.86 & 0.04 \\ 
     208 & 16.42 & 18.63 & 15.582 & 0.027 & 17.474 & 0.058 & 6.17 & 3.683 & 
0.004 & 2.65 & 0.11 & 1.69 & 0.05 \\ 
     209 & 16.42 & 18.39 & 15.578 & 0.027 & 17.470 & 0.058 & 8.39 & 3.657 & 
0.007 & 2.62 & 0.11 & 1.60 & 0.05 \\ 
     210 & 15.55 & 18.36 & 15.736 & 0.027 & 17.868 & 0.202 & 6.30 & 3.681 & 
0.006 & 3.00 & 0.36 & 1.76 & 0.15 \\ 
     211 & 15.47 & 18.68 & 16.367 & 0.083 & 18.105 & 0.202 & 5.70 & 3.688 & 
0.004 & 2.43 & 0.39 & 1.30 & 0.16 \\ 
     212 & 15.39 & 19.06 & 16.023 & 0.061 & 17.959 & 0.202 & 3.62 & 3.708 & 
0.004 & 2.74 & 0.38 & 1.62 & 0.15 \\ 
     213 & 15.15 & 18.58 & 14.538 & 0.015 & 16.357 & 0.047 & 7.22 & 3.671 & 
0.004 & 2.53 & 0.09 & 2.02 & 0.04 \\ 
     214 & 14.61 & 18.50 & 15.400 & 0.027 & 17.426 & 0.058 & 7.51 & 3.667 & 
0.008 & 2.83 & 0.11 & 1.79 & 0.05 \\ 
     215 & 14.61 & 18.04 & 12.841 & 0.005 & 14.809 & 0.006 & 5.63 & 3.689 & 
0.003 & 2.77 & 0.01 & 2.85 & 0.01 \\ 
     216 & 15.07 & 18.09 & 15.299 & 0.027 & 17.248 & 0.048 & 6.82 & 3.675 & 
0.006 & 2.73 & 0.10 & 1.81 & 0.04 \\ 
     217 & 15.61 & 17.90 & 15.055 & 0.021 & 16.851 & 0.048 & 4.29 & 3.702 & 
0.003 & 2.53 & 0.09 & 1.90 & 0.04 \\ 
     218 & 15.82 & 17.93 & 15.329 & 0.027 & 17.293 & 0.058 & 5.46 & 3.690 & 
0.004 & 2.76 & 0.11 & 1.86 & 0.05 \\ 
     219 & 16.09 & 18.01 & 15.249 & 0.021 & 17.087 & 0.048 & 7.03 & 3.673 & 
0.004 & 2.56 & 0.09 & 1.76 & 0.04 \\ 
     220 & 16.36 & 17.82 & 16.300 & 0.083 & 18.281 & 0.204 & 8.44 & 3.656 & 
0.006 & 2.75 & 0.39 & 1.36 & 0.16 \\ 
     221 & 16.96 & 17.98 & 13.942 & 0.008 & 15.841 & 0.041 & 9.37 & 3.644 & 
0.004 & 2.61 & 0.07 & 2.22 & 0.03 \\ 
     222 & 17.60 & 18.09 & 15.513 & 0.027 & 17.565 & 0.058 & 9.45 & 3.643 & 
0.006 & 2.83 & 0.11 & 1.67 & 0.05 \\ 
     223 & 17.12 & 17.74 & 14.996 & 0.021 & 16.761 & 0.048 & 8.00 & 3.661 & 
0.004 & 2.44 & 0.09 & 1.78 & 0.04 \\ 
     224 & 16.98 & 17.58 & 13.668 & 0.005 & 15.518 & 0.007 & 9.04 & 3.648 & 
0.004 & 2.54 & 0.02 & 2.31 & 0.01 \\ 
     225 & 16.52 & 17.55 & 15.812 & 0.061 & 18.124 & 0.202 & 5.87 & 3.686 & 
0.004 & 3.28 & 0.38 & 1.85 & 0.15 \\ 
     226 & 16.31 & 17.28 & 14.495 & 0.015 & 16.480 & 0.047 & 9.03 & 3.648 & 
0.004 & 2.74 & 0.09 & 2.06 & 0.04 \\ 
     227 & 15.74 & 17.39 & 15.275 & 0.027 & 17.219 & 0.048 & 6.77 & 3.676 & 
0.006 & 2.72 & 0.10 & 1.82 & 0.04 \\ 
     228 & 15.34 & 17.39 & 15.470 & 0.027 & 17.529 & 0.058 & 6.90 & 3.675 & 
0.005 & 2.89 & 0.11 & 1.80 & 0.05 \\ 
     229 & 15.42 & 16.98 & 16.262 & 0.083 & 18.066 & 0.202 & 6.45 & 3.680 & 
0.007 & 2.52 & 0.39 & 1.35 & 0.16 \\ 
     230 & 15.34 & 16.71 & 15.123 & 0.021 & 17.461 & 0.058 & 6.69 & 3.677 & 
0.005 & 3.30 & 0.11 & 2.12 & 0.05 \\ 
     231 & 15.93 & 16.71 & 13.527 & 0.005 & 15.422 & 0.007 & 10.90 & 3.622 & 
0.006 & 2.57 & 0.02 & 2.31 & 0.02 \\ 
     232 & 16.44 & 17.15 & 13.639 & 0.005 & 15.636 & 0.007 & 10.57 & 3.627 & 
0.005 & 2.73 & 0.02 & 2.34 & 0.02 \\ 
     233 & 16.98 & 17.17 & 14.927 & 0.021 & 16.440 & 0.047 & 7.26 & 3.670 & 
0.004 & 2.07 & 0.09 & 1.68 & 0.04 \\ 
     234 & 17.25 & 17.04 & 15.273 & 0.027 & 17.212 & 0.048 & 8.20 & 3.659 & 
0.006 & 2.69 & 0.10 & 1.76 & 0.04 \\ 
     235 & 17.39 & 17.20 & 14.481 & 0.015 & 16.331 & 0.047 & 8.79 & 3.651 & 
0.005 & 2.55 & 0.09 & 2.00 & 0.04 \\ 
     236 & 17.77 & 16.69 & 14.440 & 0.015 & 16.309 & 0.047 & 7.90 & 3.663 & 
0.007 & 2.59 & 0.09 & 2.06 & 0.04 \\ 
     237 & 17.31 & 16.69 & 15.226 & 0.021 & 16.971 & 0.048 & 4.79 & 3.697 & 
0.004 & 2.45 & 0.09 & 1.79 & 0.04 \\ 
     238 & 16.98 & 16.77 & 16.383 & 0.083 & 18.381 & 0.204 & 6.51 & 3.679 & 
0.008 & 2.80 & 0.39 & 1.42 & 0.16 \\ 
     239 & 17.52 & 16.07 & 15.026 & 0.021 & 16.916 & 0.048 & 10.57 & 3.627 & 
0.009 & 2.57 & 0.09 & 1.72 & 0.05 \\ 
     240 & 17.31 & 15.71 & 13.073 & 0.005 & 15.095 & 0.006 & 10.61 & 3.626 & 
0.005 & 2.77 & 0.02 & 2.58 & 0.02 \\ 
     241 & 16.39 & 15.63 & 15.333 & 0.027 & 17.192 & 0.048 & 6.66 & 3.677 & 
0.005 & 2.59 & 0.10 & 1.75 & 0.04 \\ 
     242 & 16.12 & 15.42 & 14.285 & 0.015 & 16.163 & 0.041 & 8.51 & 3.655 & 
0.008 & 2.59 & 0.08 & 2.10 & 0.04 \\ 
     243 & 16.04 & 15.77 & 15.011 & 0.021 & 16.470 & 0.047 & 4.92 & 3.696 & 
0.004 & 2.02 & 0.09 & 1.70 & 0.04 \\ 
     244 & 15.28 & 15.47 & 15.066 & 0.021 & 16.950 & 0.048 & 6.97 & 3.674 & 
0.005 & 2.63 & 0.09 & 1.86 & 0.04 \\ 
     245 & 15.31 & 16.09 & 15.352 & 0.027 & 17.243 & 0.048 & 6.55 & 3.679 & 
0.007 & 2.64 & 0.10 & 1.76 & 0.05 \\ 
     246 & 15.50 & 16.42 & 15.382 & 0.027 & 17.339 & 0.058 & 7.03 & 3.673 & 
0.004 & 2.73 & 0.11 & 1.77 & 0.05 \\ 
     247 & 15.12 & 16.36 & 14.726 & 0.015 & 16.606 & 0.047 & 10.30 & 3.631 & 
0.006 & 2.56 & 0.09 & 1.85 & 0.04 \\ 
     248 & 18.71 & 13.90 & 14.621 & 0.016 & 17.520 & 0.065 & 11.38 & 3.615 & 
0.005 & 4.04 & 0.12 & 2.44 & 0.05 \\ 
     249 & 18.58 & 13.80 & 13.958 & 0.009 & 15.403 & 0.009 & 6.33 & 3.681 & 
0.004 & 1.99 & 0.02 & 2.07 & 0.02 \\ 
     250 & 17.87 & 13.93 & 14.910 & 0.027 & 16.616 & 0.044 & 4.50 & 3.700 & 
0.004 & 2.39 & 0.09 & 1.90 & 0.04 \\ 
     251 & 16.98 & 13.61 & 14.856 & 0.027 & 17.116 & 0.055 & 8.54 & 3.655 & 
0.004 & 3.16 & 0.11 & 2.10 & 0.05 \\ 
     252 & 16.17 & 13.66 & 16.296 & 0.100 & 18.025 & 0.198 & 5.75 & 3.687 & 
0.006 & 2.41 & 0.39 & 1.32 & 0.16 \\ 
     253 & 15.93 & 13.45 & 15.774 & 0.064 & 17.581 & 0.065 & 5.53 & 3.690 & 
0.007 & 2.53 & 0.16 & 1.58 & 0.07 \\ 
     254 & 16.33 & 13.23 & 14.934 & 0.027 & 16.765 & 0.055 & 6.30 & 3.681 & 
0.004 & 2.56 & 0.11 & 1.90 & 0.05 \\ 
     255 & 16.60 & 13.45 & 16.083 & 0.064 & 18.269 & 0.233 & 4.92 & 3.696 & 
0.003 & 3.10 & 0.43 & 1.70 & 0.17 \\ 
     256 & 16.90 & 13.45 & 15.023 & 0.027 & 16.766 & 0.055 & 6.39 & 3.680 & 
0.004 & 2.43 & 0.11 & 1.81 & 0.05 \\ 
     257 & 17.09 & 13.18 & 15.868 & 0.064 & 17.650 & 0.065 & 5.28 & 3.692 & 
0.006 & 2.50 & 0.16 & 1.54 & 0.07 \\ 
     258 & 17.28 & 13.18 & 14.923 & 0.027 & 16.713 & 0.044 & 6.56 & 3.678 & 
0.004 & 2.49 & 0.09 & 1.88 & 0.04 \\ 
     259 & 17.55 & 13.31 & 15.131 & 0.027 & 16.904 & 0.055 & 7.08 & 3.672 & 
0.005 & 2.46 & 0.11 & 1.76 & 0.05 \\ 
     260 & 18.14 & 13.58 & 15.066 & 0.027 & 16.834 & 0.055 & 7.71 & 3.665 & 
0.004 & 2.45 & 0.11 & 1.76 & 0.05 \\ 
     261 & 18.31 & 13.50 & 15.182 & 0.027 & 16.984 & 0.055 & 7.14 & 3.672 & 
0.004 & 2.50 & 0.11 & 1.76 & 0.05 \\ 
     262 & 18.90 & 13.20 & 11.054 & 0.002 & 12.618 & 0.002 & 11.69 & 3.610 & 
0.005 & 2.06 & 0.01 & 3.06 & 0.01 \\ 
     263 & 18.50 & 12.58 & 14.975 & 0.027 & 16.649 & 0.044 & 7.40 & 3.669 & 
0.005 & 2.31 & 0.09 & 1.75 & 0.04 \\ 
     264 & 18.31 & 12.28 & 12.522 & 0.003 & 14.414 & 0.005 & 10.74 & 3.624 & 
0.004 & 2.57 & 0.01 & 2.72 & 0.01 \\ 
     265 & 18.71 & 12.26 & 15.774 & 0.064 & 17.527 & 0.065 & 5.80 & 3.687 & 
0.005 & 2.45 & 0.16 & 1.54 & 0.07 \\ 
     266 & 19.17 & 12.18 & 12.940 & 0.005 & 14.683 & 0.005 & 11.04 & 3.620 & 
0.007 & 2.34 & 0.02 & 2.45 & 0.02 \\ 
     267 & 17.87 & 12.64 & 15.189 & 0.027 & 17.110 & 0.055 & 7.94 & 3.662 & 
0.005 & 2.67 & 0.11 & 1.79 & 0.05 \\ 
     268 & 17.74 & 12.26 & 13.417 & 0.008 & 15.210 & 0.006 & 9.23 & 3.646 & 
0.004 & 2.46 & 0.02 & 2.37 & 0.01 \\ 
     269 & 17.15 & 12.18 & 16.087 & 0.064 & 18.519 & 0.233 & 6.83 & 3.675 & 
0.006 & 3.44 & 0.43 & 1.78 & 0.17 \\ 
     270 & 16.98 & 12.31 & 15.043 & 0.027 & 16.845 & 0.055 & 8.59 & 3.654 & 
0.003 & 2.48 & 0.11 & 1.75 & 0.05 \\ 
     271 & 16.69 & 12.18 & 12.248 & 0.003 & 14.120 & 0.003 & 12.93 & 3.591 & 
0.006 & 2.48 & 0.02 & 2.70 & 0.02 \\ 
     272 & 16.77 & 12.99 & 14.951 & 0.027 & 16.704 & 0.044 & 6.96 & 3.674 & 
0.003 & 2.43 & 0.09 & 1.83 & 0.04 \\ 
     273 & 16.39 & 12.61 & 12.552 & 0.003 & 14.315 & 0.005 & 4.08 & 3.704 & 
0.002 & 2.48 & 0.01 & 2.89 & 0.01 \\ 
     274 & 15.55 & 13.10 & 15.324 & 0.039 & 17.308 & 0.065 & 4.81 & 3.697 & 
0.003 & 2.80 & 0.13 & 1.89 & 0.06 \\ 
     275 & 15.42 & 12.50 & 14.718 & 0.016 & 16.613 & 0.044 & 6.34 & 3.681 & 
0.003 & 2.65 & 0.08 & 2.03 & 0.03 \\ 
     276 & 15.96 & 12.23 & 13.624 & 0.008 & 15.550 & 0.009 & 8.26 & 3.658 & 
0.004 & 2.67 & 0.02 & 2.41 & 0.01 \\ 
     277 & 15.47 & 11.77 & 15.327 & 0.039 & 17.333 & 0.065 & 4.94 & 3.696 & 
0.005 & 2.83 & 0.13 & 1.90 & 0.06 \\ 
     278 & 15.50 & 11.56 & 15.479 & 0.039 & 17.545 & 0.065 & 6.01 & 3.684 & 
0.008 & 2.91 & 0.14 & 1.84 & 0.06 \\ 
     279 & 15.85 & 11.34 & 13.934 & 0.009 & 15.885 & 0.045 & 7.63 & 3.666 & 
0.006 & 2.72 & 0.08 & 2.32 & 0.04 \\ 
     280 & 16.01 & 11.15 & 15.087 & 0.027 & 17.174 & 0.055 & 9.49 & 3.642 & 
0.008 & 2.89 & 0.11 & 1.86 & 0.05 \\ 
     281 & 15.82 & 11.10 & 15.398 & 0.039 & 17.480 & 0.065 & 6.14 & 3.683 & 
0.005 & 2.93 & 0.13 & 1.87 & 0.06 \\ 
     282 & 16.50 & 11.58 & 15.047 & 0.027 & 16.877 & 0.055 & 8.00 & 3.661 & 
0.010 & 2.53 & 0.11 & 1.79 & 0.05 \\ 
     283 & 16.74 & 11.58 & 14.931 & 0.027 & 16.922 & 0.055 & 8.07 & 3.661 & 
0.009 & 2.77 & 0.11 & 1.93 & 0.05 \\ 
     284 & 16.71 & 11.31 & 15.040 & 0.027 & 17.024 & 0.055 & 9.80 & 3.638 & 
0.005 & 2.73 & 0.11 & 1.81 & 0.05 \\ 
     285 & 16.71 & 10.99 & 15.251 & 0.039 & 17.147 & 0.055 & 10.42 & 3.629 & 
0.009 & 2.58 & 0.12 & 1.64 & 0.06 \\ 
     286 & 16.96 & 10.66 & 16.019 & 0.064 & 17.658 & 0.065 & 5.89 & 3.686 & 
0.006 & 2.28 & 0.16 & 1.37 & 0.07 \\ 
     287 & 17.17 & 11.20 & 15.337 & 0.039 & 17.459 & 0.065 & 8.96 & 3.649 & 
0.007 & 2.95 & 0.14 & 1.81 & 0.06 \\ 
     288 & 17.20 & 11.58 & 15.334 & 0.039 & 17.229 & 0.055 & 5.43 & 3.691 & 
0.005 & 2.66 & 0.12 & 1.81 & 0.05 \\ 
     289 & 17.58 & 11.29 & 15.161 & 0.027 & 17.013 & 0.055 & 4.87 & 3.696 & 
0.002 & 2.61 & 0.11 & 1.88 & 0.05 \\ 
     290 & 18.28 & 11.50 & 15.082 & 0.027 & 16.835 & 0.055 & 6.09 & 3.684 & 
0.005 & 2.44 & 0.11 & 1.81 & 0.05 \\ 
     291 & 18.33 & 11.69 & 15.143 & 0.027 & 16.849 & 0.055 & 5.51 & 3.690 & 
0.004 & 2.38 & 0.11 & 1.78 & 0.05 \\ 
     292 & 18.06 & 10.77 & 14.916 & 0.027 & 16.814 & 0.055 & 10.69 & 3.625 & 
0.008 & 2.58 & 0.11 & 1.76 & 0.05 \\ 
     293 & 17.77 & 10.75 & 14.598 & 0.016 & 16.558 & 0.044 & 11.05 & 3.620 & 
0.005 & 2.66 & 0.08 & 1.91 & 0.04 \\ 
     294 & 17.63 & 10.34 & 15.301 & 0.039 & 17.319 & 0.065 & 8.23 & 3.659 & 
0.007 & 2.81 & 0.14 & 1.79 & 0.06 \\ 
     295 & -6.67 & 18.85 & 14.814 & 0.014 & 16.883 & 0.042 & 10.09 & 3.634 & 
0.007 & 2.85 & 0.08 & 1.93 & 0.04 \\ 
     296 & -7.51 & 19.31 & 15.600 & 0.029 & 17.609 & 0.051 & 4.45 & 3.700 & 
0.004 & 2.84 & 0.11 & 1.81 & 0.05 \\ 
     297 & -7.70 & 19.12 & 15.219 & 0.014 & 17.213 & 0.042 & 5.71 & 3.688 & 
0.004 & 2.81 & 0.08 & 1.91 & 0.03 \\ 
     298 & -8.29 & 19.01 & 14.783 & 0.014 & 16.753 & 0.042 & 8.22 & 3.659 & 
0.005 & 2.74 & 0.08 & 1.97 & 0.03 \\ 
     299 & -8.37 & 18.82 & 15.067 & 0.014 & 16.961 & 0.042 & 5.94 & 3.685 & 
0.003 & 2.66 & 0.08 & 1.90 & 0.03 \\ 
     300 & -9.15 & 18.74 & 11.437 & 0.002 & 13.493 & 0.002 & 8.46 & 3.656 & 
0.003 & 2.86 & 0.01 & 3.35 & 0.01 \\ 
     301 & -8.26 & 18.23 & 14.468 & 0.011 & 16.434 & 0.044 & 9.28 & 3.645 & 
0.004 & 2.71 & 0.08 & 2.05 & 0.03 \\ 
     302 & -6.72 & 18.14 & 13.651 & 0.005 & 15.524 & 0.007 & 9.17 & 3.646 & 
0.004 & 2.58 & 0.02 & 2.33 & 0.01 \\ 
     303 & -7.07 & 17.77 & 15.142 & 0.014 & 17.064 & 0.042 & 7.86 & 3.663 & 
0.007 & 2.67 & 0.08 & 1.82 & 0.04 \\ 
     304 & -8.13 & 17.63 & 15.598 & 0.029 & 17.594 & 0.051 & 5.03 & 3.695 & 
0.004 & 2.82 & 0.11 & 1.78 & 0.05 \\ 
     305 & -8.48 & 17.74 & 15.391 & 0.029 & 17.322 & 0.051 & 7.87 & 3.663 & 
0.005 & 2.68 & 0.11 & 1.72 & 0.05 \\ 
     306 & -8.72 & 17.63 & 15.467 & 0.029 & 17.360 & 0.051 & 7.25 & 3.670 & 
0.004 & 2.64 & 0.11 & 1.69 & 0.05 \\ 
     307 & -9.94 & 17.90 & 15.508 & 0.029 & 17.631 & 0.051 & 10.16 & 3.633 & 
0.006 & 2.93 & 0.11 & 1.69 & 0.05 \\ 
     308 & -10.07 & 17.28 & 15.566 & 0.029 & 17.617 & 0.051 & 7.43 & 3.668 & 
0.006 & 2.87 & 0.11 & 1.74 & 0.05 \\ 
     309 & -10.12 & 17.50 & 15.794 & 0.060 & 17.879 & 0.196 & 5.97 & 3.685 & 
0.007 & 2.94 & 0.36 & 1.72 & 0.15 \\ 
     310 & -8.88 & 17.28 & 15.160 & 0.014 & 17.095 & 0.042 & 8.64 & 3.653 & 
0.005 & 2.68 & 0.08 & 1.78 & 0.03 \\ 
     311 & -8.80 & 17.15 & 15.482 & 0.029 & 17.506 & 0.051 & 9.00 & 3.649 & 
0.007 & 2.80 & 0.11 & 1.69 & 0.05 \\ 
     312 & -8.56 & 17.31 & 15.606 & 0.029 & 17.457 & 0.051 & 8.29 & 3.658 & 
0.004 & 2.56 & 0.11 & 1.57 & 0.05 \\ 
     313 & -8.05 & 16.98 & 15.477 & 0.029 & 17.505 & 0.051 & 7.95 & 3.662 & 
0.006 & 2.83 & 0.11 & 1.74 & 0.05 \\ 
     314 & -6.94 & 17.12 & 15.234 & 0.014 & 17.284 & 0.051 & 9.04 & 3.648 & 
0.007 & 2.84 & 0.10 & 1.80 & 0.04 \\ 
     315 & -6.59 & 16.98 & 15.524 & 0.029 & 17.516 & 0.051 & 5.68 & 3.688 & 
0.005 & 2.80 & 0.11 & 1.79 & 0.05 \\ 
     316 & -6.86 & 16.71 & 15.647 & 0.029 & 17.546 & 0.051 & 6.88 & 3.675 & 
0.004 & 2.65 & 0.11 & 1.64 & 0.05 \\ 
     317 & -6.94 & 16.58 & 15.254 & 0.029 & 17.181 & 0.042 & 6.51 & 3.679 & 
0.003 & 2.70 & 0.09 & 1.83 & 0.04 \\ 
     318 & -7.64 & 16.39 & 15.395 & 0.029 & 17.375 & 0.051 & 8.66 & 3.653 & 
0.007 & 2.74 & 0.11 & 1.72 & 0.05 \\ 
     319 & -6.61 & 15.77 & 14.757 & 0.014 & 16.307 & 0.044 & 10.12 & 3.633 & 
0.006 & 2.08 & 0.08 & 1.65 & 0.04 \\ 
     320 & -7.48 & 15.36 & 15.453 & 0.029 & 17.392 & 0.051 & 8.26 & 3.658 & 
0.005 & 2.69 & 0.11 & 1.68 & 0.05 \\ 
     321 & -7.75 & 15.55 & 13.625 & 0.005 & 15.630 & 0.007 & 10.90 & 3.622 & 
0.007 & 2.73 & 0.02 & 2.33 & 0.02 \\ 
     322 & -8.34 & 16.52 & 13.477 & 0.005 & 15.438 & 0.007 & 8.02 & 3.661 & 
0.004 & 2.73 & 0.02 & 2.50 & 0.01 \\ 
     323 & -8.96 & 16.96 & 14.418 & 0.011 & 16.562 & 0.044 & 9.48 & 3.642 & 
0.005 & 2.97 & 0.08 & 2.17 & 0.04 \\ 
     324 & -9.10 & 16.79 & 13.835 & 0.008 & 15.898 & 0.039 & 10.18 & 3.633 & 
0.005 & 2.84 & 0.07 & 2.32 & 0.03 \\ 
     325 & -8.53 & 15.66 & 12.615 & 0.003 & 14.755 & 0.006 & 9.82 & 3.638 & 
0.005 & 2.96 & 0.01 & 2.87 & 0.02 \\ 
     326 & -8.59 & 15.34 & 13.698 & 0.005 & 15.683 & 0.007 & 10.55 & 3.627 & 
0.005 & 2.71 & 0.02 & 2.31 & 0.01 \\ 
     327 & -9.05 & 15.28 & 12.764 & 0.004 & 14.750 & 0.005 & 9.86 & 3.637 & 
0.006 & 2.73 & 0.02 & 2.72 & 0.02 \\ 
     328 & -9.18 & 15.85 & 15.145 & 0.014 & 17.151 & 0.042 & 6.83 & 3.675 & 
0.004 & 2.81 & 0.08 & 1.90 & 0.03 \\ 
     329 & -9.99 & 15.98 & 14.581 & 0.011 & 16.522 & 0.044 & 5.01 & 3.695 & 
0.004 & 2.73 & 0.08 & 2.16 & 0.03 \\ 
\label{gc_table}
\end{longtable*}

\subsection{Observed Hertzsprung-Russell Diagram}

\begin{figure*}
\epsscale{1.0}
\includegraphics[width=.8\textwidth,angle=90]{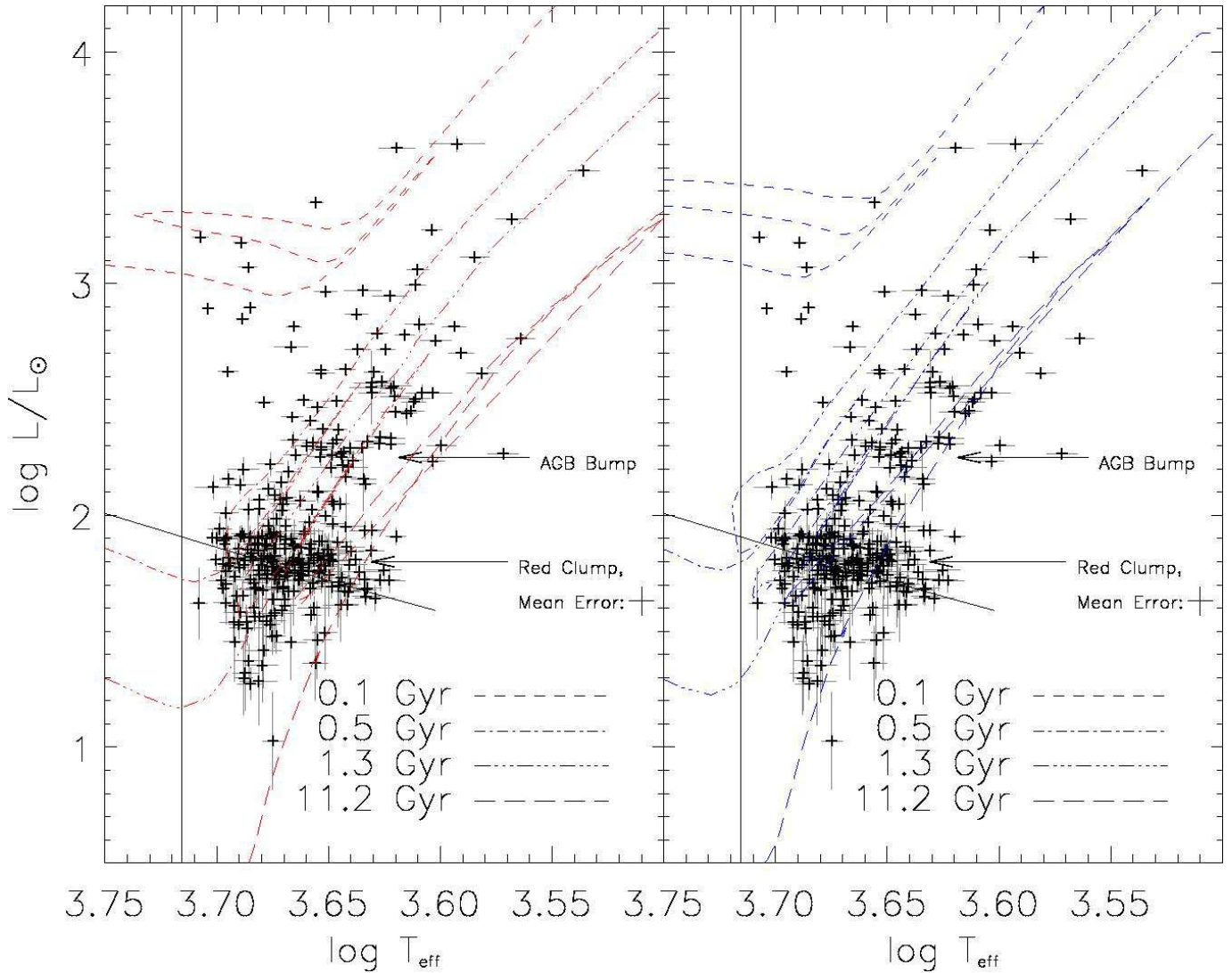}
\caption{H-R diagram for the GC stars with solar-metallicity (in red,
left) and metal-poor (Z=0.008, in blue, right) isochrones from
\citet{Girardi00} overplotted. The magnitude limit for 50\%
completeness (see Figure \ref{comp_err}) and the temperature limit for
which stars no longer show CO (see Figure \ref{comp_relation}) are
overplotted as solid dark lines.  Errors are shown in grey.  The mean
errors in temperature and luminosity for stars with luminosities $1.6
< \log(L/ L_{\sun}) < 2.0$ (the red clump region) are shown at right.}
\label{hr}
\end{figure*}

\begin{figure}
\epsscale{1.0}
\includegraphics[width=.35\textwidth,angle=90]{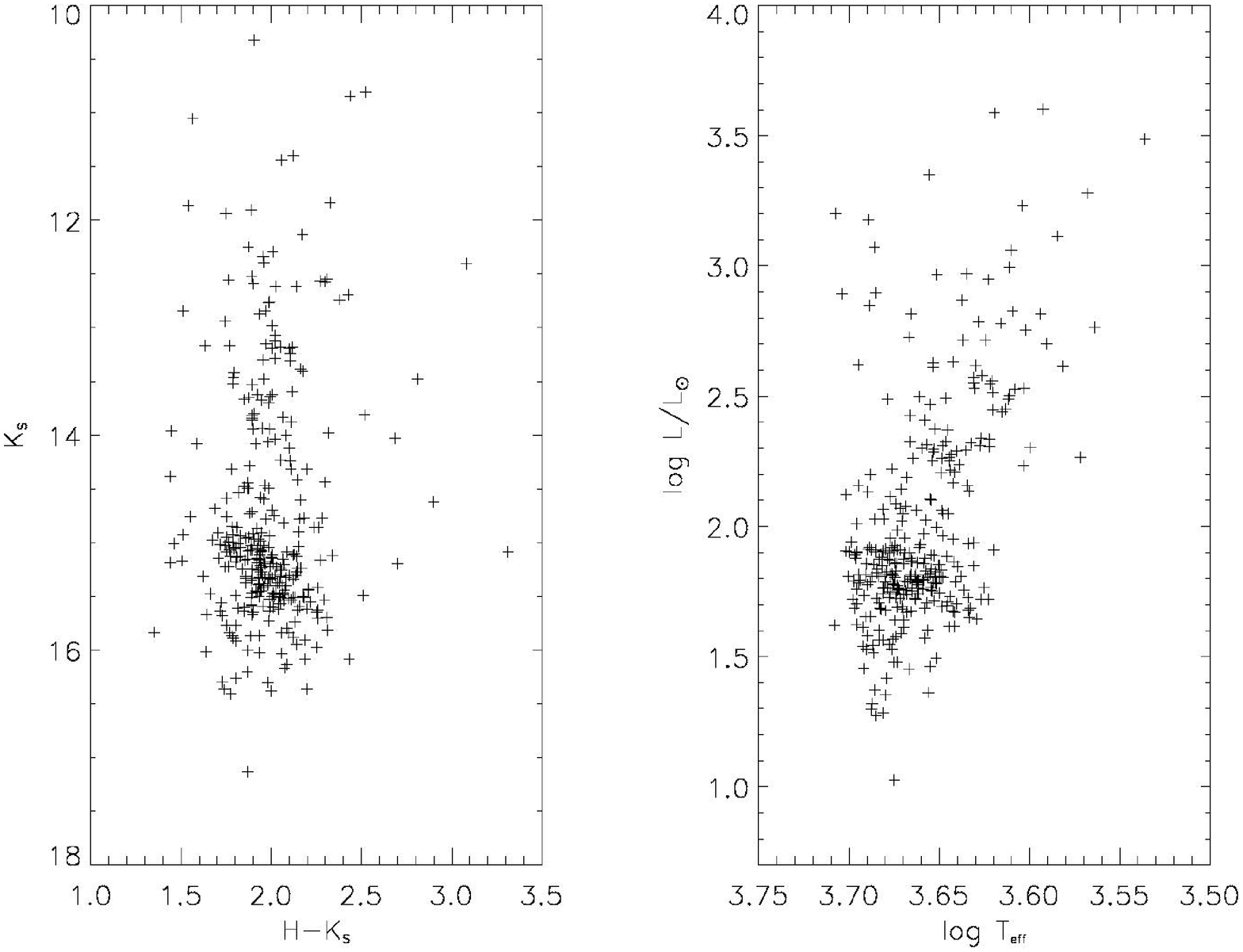}
\caption{Comparison of the observed GC H-R diagram to a
color-magntidue diagram for the same GC stars.  The figure
demonstrates the advantange of the spectroscopic temperature
derivation.  The RGB/AGB and RC populations are more clearly
distinguished in the H-R diagram than in the CMD, and there is
less scatter in \teff than $H - K$, due to variations in GC
extinction.}
\label{cmd}
\end{figure}

The GC H-R diagram is shown in Figure \ref{hr}.  Our data clearly show
the red clump at $\log(L/ L_{\sun}) \sim 1.7$, as well as the upper
red giant branch / early asymptotic giant branch.  There is an
indication of the AGB bump at $\log(L/L_{\sun}) \sim 2.3$, and we also
detect some lower red giant branch stars at lower luminosities, though
the observations are highly incomplete in this region.  The data set
in Figure \ref{hr} allows for a more robust analysis of the Galactic
Center star formation history than any other previously published data
set.  Photometric studies are limited to the modeling of luminosities
alone \citep{Rieke87,Narayanan96, Davidge97, Philipp99, Alexander99,
Figer04}, due to intrinsic variations in late-type giant colors and
the large variation in Galactic Center extinction.  This point is
illustrated in Figure \ref{cmd}, which compares the observed H-R
diagram to a color-magnitude diagram (CMD) for the same GC stars.  The
RGB/AGB and RC populations are more clearly distinguished in the H-R
diagram than in the CMD, and there is less scatter in \teff than $H -
K$, due to variations in GC extinction.  As a result, only the K-band
luminosity function can reliably be modeled with broadband photometry
alone.  The observed H-R diagram in this study is also an improvement
with respect to previous spectroscopic work \citep{Blum96b,Blum03},
due to our improved magnitude limit ($\sim$5 mag. deeper), which
allows for the analysis of well-populated regions in the H-R diagram,
in which the evolutionary models are fairly well understood.

The errors in temperature and luminosity are shown with the data in
grey in Figure \ref{hr}.  Average errors in temperature and luminosity
for stars with luminosities $1.6 < \log(L/ L_{\sun}) < 2.0$ (the red
clump region) are shown at right. The temperature limit for the
minimum CO index and the 50\% completeness limit for the average
extinction correction ($A_K = 2.75$ mag.) are also shown.  Solar
metallicity (in red, left) and metal-poor (Z=0.008, in blue, right)
isochrones from \citet{Girardi00} are overplotted.  The isochrones
show the wide range of ages spanned by the Galactic Center population.
Figure \ref{hr} also demonstrates the age-metallicity degeneracy
inherent in this part of the diagram.  Stars in the same part of the
H-R diagram may represent a metal-rich, intermediate-age ($\lesssim$ a
few Gyr) population or a metal-poor, old ($\gtrsim$ 5 Gyr) population.
For stars younger than $\sim$5 Gyr, the Galactic Center stellar
population is known to be approximately solar
\citep{Ramirez00,Carr00}.  For older stars, the Galactic Center
metallicity distribution is not well-known.  We will address this
uncertainty later in the paper.

\subsection{Deriving the Star Formation History}

To investigate the star formation history implied by our sample, we
begin with a qualitative discussion, followed by quantitative analysis
in \S 3.4-3.5.  We start by considering three models based on
candidate star formation histories presented in the literature. The
first scenario consists of an ancient burst of star formation 7.5-8.5
Gyr ago, similar to the single stellar population of the bulge.
\citet{Genzel03} compared this scenario to the K band luminosity
function (KLF) of the inner parsec.  The second model consists of
constant star formation between 10 Myr and 10 Gyr ago.  This model is
based on the best-fit model of \citet{Figer04}, who considered the
Galactic Center KLF within 40 pc of the center.  The third model
corresponds to the best-fit star formation history of \citet{Blum03},
who fit the H-R diagram of asymptotic giant branch and cool supergiant
stars within the central 5 pc, using four specified age bins.  All
models are summarized in Table \ref{sf_models}.

\begin{deluxetable*}{llccccccccc}
\tabletypesize{\scriptsize}
\tablecaption{Summary of Star Formation History Models}
\tablecolumns{11}
\tablewidth{0pt}
\tablehead{
\colhead{Model} & \colhead{Description} & \colhead{Age} & \colhead{Z} &
\colhead{IMF Slope} & \colhead{$M_{lower}$} & \colhead{$M_{upper}$} &
\colhead{$\chi_{\lambda}^2$\tablenotemark{a}} & \colhead{$P_{\lambda}$\tablenotemark{b}} & 
\colhead{Relative SFR\tablenotemark{c}} & \colhead{$\sigma_{SFR}$\tablenotemark{d}} \\
 & & (Gyr) & & & ($M_{\sun}$) & ($M_{\sun}$) & &  (\%) & &
}
\startdata
1  &  Bulge-like       &  7.50 - 8.50  &  0.015         &  2.35  &  0.7  &  120  & 3228.8 & 0.00  & 1.00 & -     \\ 
2  &  Continuous       &  0.01 - 10.0  &  0.019         &  2.35  &  0.7  &  120  & 404.9  & 3.76  & 1.00 & -     \\ 
3  &  Blum et al.      &  0.01 - 0.10  &  0.019         &  2.35  &  0.7  &  120  & 678.8  & 0.01  & 0.65 & -     \\ 
   &                   &  0.10 - 1.00  &  0.019         &  2.35  &  0.7  &  120  &        &       & 0.06 & -	 \\ 
   &                   &  1.00 - 5.00  &  0.019         &  2.35  &  0.7  &  120  &        &       & 0.09 & -	 \\ 
   &                   &  5.00 - 12.0  &  0.019         &  2.35  &  0.7  &  120  &        &       & 0.21 & -	 \\ 
4  &  Two-bin, solar   &  0.01 - 5.00  &  0.019         &  2.35  &  0.7  &  120  & 384.5  & 3.03  & 0.82 & 0.04  \\ 
   &                   &  5.00 - 12.0  &  0.019         &  2.35  &  0.7  &  120  &        &       & 0.18 & 0.07	 \\  
5  &  Two-bin, solar   &  0.01 - 7.00  &  0.019         &  2.35  &  0.7  &  120  & 270.0  & 16.66 & 0.90 & 0.04  \\ 
   &                   &  7.00 - 12.0  &  0.019         &  2.35  &  0.7  &  120  &        &       & 0.10 & 0.09	 \\ 
6  &  Two-bin, poor    &  0.01 - 5.00  &  0.019         &  2.35  &  0.7  &  120  & 450.0  & 0.68  & 0.91 & 0.05  \\ 
   &                   &  5.00 - 12.0  &  0.008         &  2.35  &  0.7  &  120  &        &       & 0.09 & 0.08  \\ 
7  &  Closed-box       &  0.01 - 12.0  &  0.004 - 0.019 &  2.35  &  0.7  &  120  & 1035.6 & 0.00  & 1.00 & -  \\ 
8  &  Flat IMF         &  0.01 - 12.0  &  0.019         &  0.85  &  0.7  &  120  & 242.8  & 39.70 & 1.00 & -  \\ 
9  &  High $M_{lower}$ &  0.01 - 12.0  &  0.019         &  2.35  &  2.5  &  120  & 4201.8 & 0.00  & 1.00 & -  \\ 
10 &  High $M_{lower}$ &  0.01 - 12.0  &  0.019         &  2.35  &  1.5  &  120  & 551.4  & 0.00  & 1.00 & -  \\ 
11 &  Model 5+8 Hybrid &  0.01 - 7.00  &  0.019         &  0.85  &  0.7  &  120  & 257.0  & 26.3  & 0.48 & 0.02 \\ 
   &                   &  7.00 - 12.0  &  0.019         &  0.85  &  0.7  &  120  &        &       & 0.51 & 0.20 \\ 
\enddata
\label{sf_models}
\tablenotetext{a}{Poisson likelihood parameter: $\chi_{\lambda}^2 =
2\sum_{i}m_{i} - n_{i} + n_{i}\ln \frac{n_i}{m_i}$; $n_{i}$ is
the number of observed stars and $m_{i}$ is the number of model stars
in bin $i$. }  
\tablenotetext{b}{Goodness of Fit Parameter: Percentage
of Monte Carlo trials in which synthetic data sets composed of stars
drawn from the best-fit models have larger $\chi_{\lambda}^2$ than the
$\chi_{\lambda}^2$ implied by the observed data set. If the model is
an accurate representation of the true star formation history, this
percentage should be approximately 50\%.}  
\tablenotetext{c}{Average
relative star formation rate for each age bin specified in the third 
column.  The rates are normalized such that the total relative star formation 
rate is 1.}
\tablenotetext{d}{Uncertainty in the relative star formation rate,
derived by fitting the star formation history to a series of synthetic
data sets consisting of stars drawn from the observed data
set.}
\end{deluxetable*}

For each scenario, we generated model H-R diagrams using the synthetic
color-magnitude diagram computation algorithm IAC-Star
\citep{Aparicio04}.  The algorithm uses a Monte Carlo approach to
compute composite stellar populations on a star-by-star basis, using a
specified set of evolutionary tracks.  The code accommodates several
additional inputs, including the initial mass function, star formation
rate function, and chemical enrichment law.  To compare the generated
models to the data, we added Gaussian noise based on the average
observed errors in luminosity and temperature.  We also randomly
removed stars from the models according to our estimated spectroscopic
completeness (Fig \ref{comp_err}).

\begin{figure*}
\epsscale{0.7}
\plotone{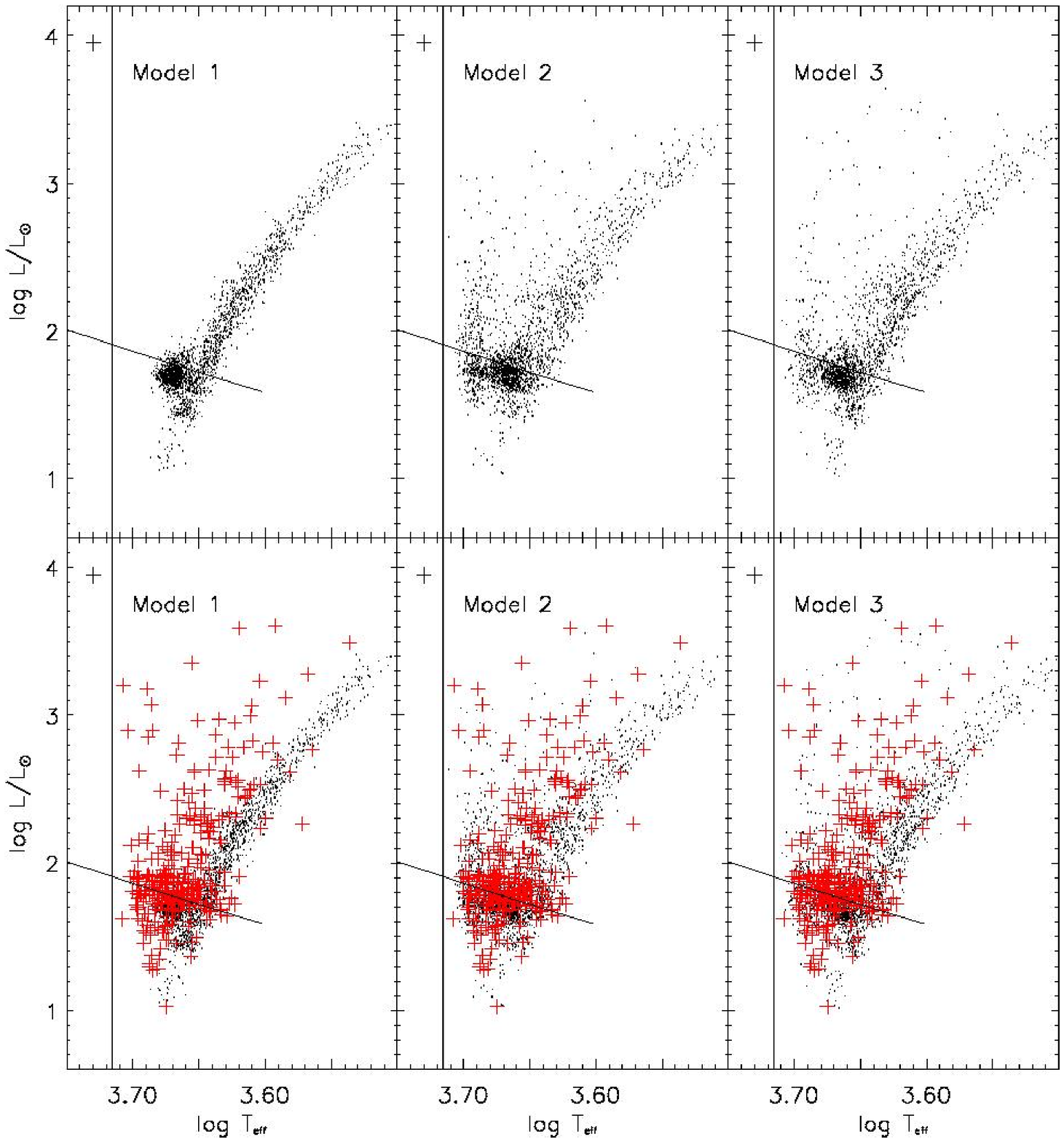}
\caption{Comparison of the observed data (red crosses) to several
proposed models given in the literature (black dots).  See the text
and Table \ref{sf_models} for a description of the models.  We have
added noise to the models based on the average observed errors in
luminosity and temperature (upper left). We also corrected for the
incompleteness function of the data (Fig \ref{comp_err}).  A
qualitative comparison suggests that all literature models produce
relatively too many cool stars to describe the observed data set.}
\label{sf_models_lit}
\end{figure*}

The models are shown against our Galactic Center sample in Figure
\ref{sf_models_lit}.  A qualitative comparison suggests that the
ancient burst model is insufficient to fully describe the observed
data set.  The continuous star formation model and the \citet{Blum03}
model span the full range in temperature and luminosity of the
observed data set but seem to have relatively too many stars at cool
temperatures.  It is unlikely this observed effect is caused by
systematic errors or a selection bias.  A systematic underestimate of
the GC CO indices would shift the pattern to hotter temperatures.
However, the red edge of the Hertzsprung gap (log \teff $\sim 3.70$)
is well matched by the models, making such a systematic effect
unlikely.  In addition, some cool RGB/AGB stars are present in the data,
while a systematic shift to warmer temperatures would allow none.
Finally, the brightness and CO strength of cool RGB/AGB stars compared to
warmer, dimmer red clump stars means that cool RGB/AGB stars are
relatively easy to detect and classify.  There is no obvious way of
selectively removing these stars from the sample.

Motivated by the comparison in Figure \ref{sf_models_lit}, we consider
three alternative scenarios that could potentially explain the
relative paucity of stars at low temperatures:
\begin{enumerate}
\item{The first and simplest possibility is that the star
formation rate was low at early times ($\gtrsim 5$ Gyr).  To test
this hypothesis, we generated two models consisting of linear
combinations of constant star formation in two specified age bins
(Models 4 and 5 in Table \ref{sf_models}).  Model 4 assumes bins of
0.01-5 Gyr and 5-12 Gyr.  Model 5 assumes bins of 0.01-7 Gyr and 7-12 Gyr.}
\item{A second possibility is that the oldest stars are metal poor.  The
previously described models assume a solar metallicity for all times.
This assumption is motivated by the work of \citet{Ramirez00}, who
found the stellar [Fe/H] abundance to be approximately solar for stars
younger than $\sim$5 Gyr.  For older stars, the Galactic Center
metallicity distribution has not been determined.  However, the bulge,
which formed 7-12 Gyr ago, is known to have a nearly solar
distribution \citep{Sadler96}.  An extremely metal poor ancient
population would, therefore, represent a completely separate
population from the bulge.  Given the current observed differences
between the bulge and the nucleus discussed in \S 1, we can not
exclude this possibility.  We, therefore, generated two models to
represent this scenario.  The first model, Model 6, consists of a
linear combination of constant star formation in two age bins with
solar metallicity for 0.01-5 Gyr and a metallicity of 0.008 for 5-12 Gyr.
The second model, Model 7, is a simple closed box model assuming
constant star formation for 0.01-12 Gyr, a starting metallicity of 0.004
and an ending metallicity of solar.}
\item{A final way to explain the relative lack of stars at cool temperatures
invokes a non-standard initial mass function (IMF).  In all above
models, we assume a Salpeter IMF with an upper mass limit of 120
$M_{\sun}$ and a lower mass limit of 0.7 $M_{\sun}$.  As stars less
massive than $\sim 0.7 M_{\sun}$ have main sequence lifetimes
comparable to the age of the Universe, we are not sensitive to stars
below this mass limit.  Recent results indicate that the current
initial mass function of the Galactic Center is flatter than a
Salpeter function or has an unusally high lower mass cut-off
\citep{Nayakshin05,Paumard06}.  Such a mass function could explain the
relative lack of low mass stars in the observed old stellar
population.  To test this hypothesis, we consider three models, all
assuming continuous star formation at solar metallicity for 0.01-12 Gyr.
The first model, Model 8, assumes a flat single power-law slope
($dN/dm=m^{-0.85}$), chosen to match the results of \citet{Paumard06}.
Models 9 and 10 both assume a standard Salpeter slope
($dN/dm=m^{-2.35}$), but with lower mass limits of 2.5 $M_{\sun}$ and
1.5 $M_{\sun}$, respectively.}
\end{enumerate}
Results stemming from these three hypotheses are tested in \S 3.5.

\subsection{Quantification of Fit}

To quantitatively compare the models described above to the data, we
adopted the numerical techniques described in \citet{Dolphin02}.  We
first binned the observed and model H-R diagrams in temperature and
luminosity for stars above our 50\% completeness threshold, using
uniform bins with size three times our average errors ($\delta \log(L/
L_{\sun}) = 0.12$, $\delta \log T_{eff} = 0.018$).  The choice of bins
in this technique is somewhat subjective.  However, tests using
different bins sizes showed that while the binning scheme does change
the fit quality, it does not significantly affect the derived star
formation rates.  This finding is in agreement with \citet{Dolphin02}.

For models with two age bins, we used the Numerical Recipes procedure
AMOEBA \citep{Press92} to search for the linear combination of models
that minimized the Poisson maximum likelihood parameter:
$\chi_{\lambda}^2 = 2\sum_{i}m_{i} - n_{i} + n_{i}\ln \frac{n_i}{m_i}$
\citep{Dolphin97}.  Here, $n_{i}$ is the number of observed stars and
$m_{i}$ is the number of model stars in bin $i$.  For models with
fixed relative star formation rates, we scaled the model distribution
to minimize $\chi_{\lambda}^2$.  To estimate the errors in the derived
star formation history, we implemented the technique described in
\citet{Blum03}.  We built a set of 100 H-R diagrams consisting of a
random sampling of 329 stars drawn from the observed H-R diagram,
allowing each observed star to be selected any number of times.  We
then re-derived the star formation history for each H-R diagram and
given model.  The resulting standard deviation in the derived star
formation rate was taken as the 1$\sigma$ uncertainty.  

We measured the fit quality through a second set of Monte Carlo
simulations. In this set, we generated 10,000 synthetic data sets
drawn from the fitted models, selecting 329 stars randomly for each
trial and re-deriving the star formation history for the selected
subset. For a model that is an accurate representation of the true
star formation history, $\chi_{\lambda}^2$ derived from the actual
data set should be comparable to $\chi_{\lambda}^2$ derived from a
typical synthetic data set.  Therefore, to establish the goodness of
fit, we calculated the percentage of trials, $P_{\lambda}$, in which
$\chi_{\lambda}^2$ was larger when fitting the synthetic data sets to
the models than when fitting the true data sample to the models.  If
the model is a good representation of the true star formation history,
$P_{\lambda}$ should be approximately 50\%.

\subsection{Model Results}

\begin{figure*}
\epsscale{1.0}
\plotone{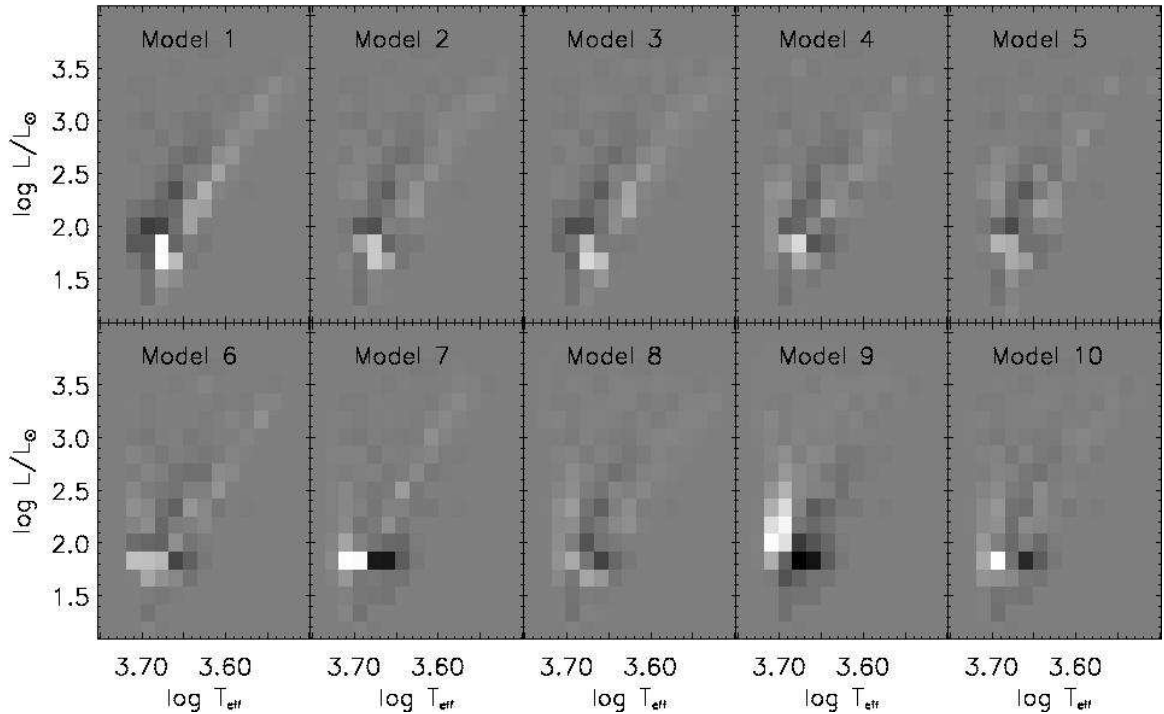}
\caption{Hess diagrams showing the difference between the observed
data histogram and the best-fit model histograms summarized in Table
\ref{sf_models}. White indicates regions where the model produces too
many stars relative to the observed data set, and black indicates
regions where the model produces too few stars relative to the
observed data set. The first three panels correspond to models
proposed in the literature and described in the text.  All show white
diagonal streaks, corresponding to the overdensity of cool stars in
the models relative to the data.  The early-history low star formation
rate models (Models 4 and 5) suffer similar systematic deficiencies.
The early-history low metallicity models (Models 6 and 7) predict a
red clump / horizontal branch morphology that is too blue compared to
the observed distribution.  The high IMF mass limit models (Models 9
and 10) predict a hotter and more luminous red clump than is observed.
The flat IMF model (Model 8) corresponds to the best fit and shows
minimal systematic trends compared to the other models.}
\label{hess}
\end{figure*}

The results of all model fits are listed in Table \ref{sf_models}.
Hess diagrams showing the difference between the observed data
histogram and the best-fit model histograms are shown in Figure
\ref{hess}.  White indicates regions where the model produces too many
stars relative to the observed data set, and black indicates regions
where the model produces too few stars relative to the observed data
set.  Examination of the $P_{\lambda}$ values in Table \ref{sf_models}
confirms the qualitative result discussed in \S 3.3 that no literature
model (Models 1-3) is a likely description of the observed data set.
This result is also reflected in the first three panels in Figure
\ref{hess}, which show white diagonal streaks, corresponding to the
overdensity of cool stars in the models relative to the data.

Table \ref{sf_models} also suggests that the early-history low star
formation rate models (Models 4 and 5) are unlikely representations of
the data, though Model 5 returns a considerably better fit, and we are
unable to completely exclude this model.  However, examination of the
fourth and fifth panels in Figure \ref{hess} show that both models
systematically overpredict the number of cool stars, as in Models 1-3.
The early-history low metallicity models (Models 6 and 7) also show
systematic deficiencies when compared to the data.  Both models
predict a red clump / horizontal branch morphology that is too blue
compared to the observed distribution.  Similar discrepencies between
the models and data are present in the high IMF mass limit models
(Models 9 and 10). Each predicts a hotter and more luminous red clump
than is observed.  However, while Models 9 and 10 predict no old, cool
stars, the flat IMF model (Model 8) predicts few old, cool stars, in
good agreement with the observations.  Model 8 also gives a
satisfactory goodness of fit ($P_{\lambda}$ = 40\%) suggesting that
this model is a likely representation of the observed distribution.

Using the models listed in Table \ref{sf_models}, we tested for
population differences as a function of distance from the cluster
center.  Such differences within the central 1-2 pc are expected due
to mass segregation, and several broadband photometry studies have
found indications of such an effect
\citep{Philipp99,Genzel03,Schodel07}.  To test for this effect in our
data, we separated the data into two sets containing the four
innermost ($4''-15''$, $11''$ median) and outermost ($13''-26''$,
$20''$ median) spectral regions shown in Figure \ref{mos}.  We then
refit the two best-fit models (Models 5 and 8) to each subset,
applying the average errors and completeness specific to that subset.
This process returned fits for each subset that are consistent with
those obtained for the entire data set.  In particular, the subset
$P_{\lambda}$ values were within 10\% of that obtained for the entire
data set, and the relative star formation rates returned by both
subsets for Model 5 were within 1$\sigma$ of the rates returned by the
entire data set.  We therefore, find no significant evidence in our
data for a population gradient.  This result is not that surprising,
as the broadband photometry results of \citet{Schodel07} suggest that
variations in the cool, low-mass stellar population within the central
parsec should be most evident 3-7 arcsec from Sgr A$^{*}$.  Our sample
is outside of the region expected to exhibit the largest population
differences (see Figure \ref{mos}).  In the future, the technique we
describe could be applied to regions closer to Sgr A$^{*}$, although
limited spectroscopic completeness due to increased stellar density
would complicate such a study.  Still, a thorough spectroscopic study
of the red-clump and RGB/AGB populations as a function of distance
from Sgr A$^{*}$ could provide the first definitive test of mass
segregation within the central parsec \citep{Alexander05}.

With the restriction to simple models, the top-heavy IMF model appears
to be superior to the remaining three model families we consider (low
star formation rate (SFR) at early times, low metallicity at early
times, IMF with high lower mass limit).  The low-SFR, low-Z, and high
$M_{lower}$ families all show consistent systematic deficiencies when
compared to the data.  It is, therefore, unlikely one could slightly
change the chosen parameters (i.e. precise metallicity, age bin
cut-offs, mass ranges, etc.)  within these families to produce a model
that is an adequate description of the data.

\section{Discussion}

Of the models considered in Table \ref{sf_models}, Model 8 (continuous
star formation at solar metallicity with a top-heavy IMF) fits the
observations best, and our Monte Carlo tests show that it is a
reasonable description of the data.  Based on the goodness of fit, we
cannot completely exclude Model 5.  However, as discussed in the
previous section, this model shows systematic deviations from the
data; we, therefore, favor Model 8.  We note that Model 8 is probably
not the only possible description of the data with this degree of
likelihood.  All models so far discussed assume a simplistic
description of the GC star formation history, and there are likely
more complicated scenarios that also adequately describe the data.
For instance, an additional model with age bins identical to Model 5
and an IMF slope identical to Model 8 yields a reasonable fit
($\chi^2_\lambda=257.0$, $P_\lambda=26.3\%$).  However, we do not feel
that increasing the number of free parameters is justified, given the
limits, uncertainties, and size of our data set.

In the context of our current knowledge about the Galactic Center,
continuous star formation with a top-heavy IMF is reasonable.  At
present, there is substantial evidence for stars in the Galactic
nucleus spanning a wide range of ages, based on results from broadband
photometry, and several distinct age tracers, including supergiants,
AGB stars, and OH/IR stars (see \S 1).  The present distribution and
kinematics of gas in the inner Galaxy is also consistent with
continuous star formation \citep{Morris96}.

In addition, the most recent epoch of star formation in the Galactic
Center is likely represented by a top-heavy initial mass function
\citep{Nayakshin05, Paumard06}.  If this recent epoch of star
formation is not anomolous and periodic bursts of GC star formation
have occurred throughout the history of the Galaxy, there is no a
priori reason to believe the GC IMF would change with time.  The
present data set appears to support this picture.  We note that the
mass traced by giants in the Central Cluster is primarily set by the
initial mass formed, rather than dynamical effects, since mass
segregation is expected to only strongly affect the central $\sim$0.01
pc of the Galactic Center \citep{Hopman06, Freitag06}.  On the 1-2 pc
scale studied in this paper, the efficiency of mass segregation is
thought to be much lower, in agreement with observational photometry
results \citep{Genzel03, Schodel07}.

Though our findings are broadly consistent with several stellar
population studies of the Galactic Center and the nucleus at large,
there remains some disagreement.  In particular, our results are
somewhat different from those presented in \citet{Blum03}, who probed
the central 5 pc and used supergiants and bright AGB stars to quantify
the star formation history in a method very similar to that employed
here.  Using their observed H-R diagram, \citet{Blum03} argued for
variable star formation over a wide range of ages, with the majority
of stars formed more than 5 Gyr ago at solar metallicity.  We find a
much smaller fraction of our sample is represented by
solar-metallicity low-mass stars with ages $\gtrsim$ 5 Gyr.  

The present work and the work of \citet{Blum03} study different
regions in the H-R diagram.  Each regime has relative advantages and
disadvantages for studying the GC star formation history, and we
discuss each in turn:

\begin{enumerate}
\item{{\it Field of View}: Due to the rarity of supergiants and bright
AGB stars in the Galactic Center, \citet{Blum03} probed all late-type
stars above their magnitude limit for the entire Central Cluster (r $<
2.5$ pc).  They, therefore, derived absolute star formation rates and
were able to compare the cluster mass implied to estimates from
dynamical studies.  In contrast, our study probes a relatively small
region on the sky ($\sim$0.2 pc$^2$ within r $< 1.0$ pc).  As
such, our results are somewhat susceptible to population
inhomogeneities in the Central Cluster, and we are able to derive only
relative star formation rates.}
\item{{\it Stellar Crowding}: Spectral extraction from our SINFONI
fields is complicated by stellar crowding.  While we took care to
ensure that our derived CO indices are largely independent of exact
pixel and background selection, contamination by neighboring stellar
spectra remains a source of uncertainty for many of the dim stars.
The stars studied by \citet{Blum03}, on the other hand, are well
separated from each other.  In addition, the GC AGBs and supergiants
are much brighter than neighboring stars, so there is negligible
uncertainty in the \citet{Blum03} spectra due to crowding.}
\item{{\it Spectral Classification}: The spectral classification
method presented by \citet{Blum03} is very similar to that presented
here.  Our depper study has the advantage that only giants are
observed, and thus, a separate luminosity class determination is not
required.  Additionally, the stars we study are warmer than those
studied by \citet{Blum03}, and thus, our CO-\teff relationship is
tighter than theirs, due to reduced uncertainties in model atmosphere
spectra at warmer temperatures \citep{Reid00}.  We further note that
the derived temperatures for the red supergiants in the work of
\citet{Blum03} are likely systematically too cool, as
\citet{Levesque05} recently showed that red supergiants are $\sim$400
K warmer than previously thought.}
\item{{\it Adequacy of Stellar Evolution Models}: The evolutionary
models in the part of the H-R diagram probed by our study (red clump,
red giant branch, early AGB) are much less uncertain than the part of
the H-R diagram studied by \citet{Blum03} (supergiants,
TP-AGBs). \citet{Gallart05} review the adequacy of AGB stellar
evolution models for deriving star formation histories, concluding
that the observed bright AGB populations are often more sensitive to
poorly known modeling parameters than the star formation history.  The
input physics for the late AGB stages is not well determined
(e.g. mass loss, convection, efficiency of the third dredge-up), and
currently, only the Padova libraries include these stages
\citep{Girardi00}.  While uncertainties in mass loss and convection
are also present for the red clump, RGB, and early AGB, the input
physics for these stages is much better understood, and a number of
stellar evolution models including these phases have been compared and
tested against observations.  For this reason, we believe our star
formation history fit is more reliable than that presented in
\citet{Blum03}.  Finally, our data set also has the benefit that it
contains several evolutionary features (red clump, AGB bump) that are
cleraly distinguishable from surrounding regions in the H-R diagram.
The diagram morphology assures us that no large systematic effects are
present in the data and allows us to construct physically motivated
models. The region of the H-R diagram studied by \citet{Blum03} has no
such morphological features.}
\end{enumerate}

While we believe that our results are more robust than those presented
in \citet{Blum03}, we also note that the findings are not necessarily
in conflict.  The region probed by \citet{Blum03} extends to $\sim$2.5
pc from the center, whereas we probe only the central parsec.
Furthermore, we note that there are some similarities in our
findings. Both studies find evidence for star formation throughout the
history of the Galaxy, and both studies suggest that purely solar
metallicity models are needed to produce the observed data.  Still,
further work is needed to resolve the remaining discrepencies (i.e.,
variable versus continuous star formation, Salpeter versus flat
IMF). Specifically, a sample tracing a large region on the sky and
containing thousands of stars would represent a significant step
forward in this field.  The planned FLAMINGOS-2 GC Survey on Gemini
\citep{Eikenberry06} will obtain 4000 late-type giant spectra out to
one degree in Galacto-centric radius ($\sim$140 pc) and with a
spectral resolution of R$\sim$20,000.  It will, therefore, provide
unprecedented information on the Galactic Center chemical enrichment
and star formation history.

\section{Summary and Conclusion}
We observed 329 late-type giants $4''-26''$ north of Sgr A$^{*}$ with
the integral field spectrometer SINFONI on the VLT.  Combining
spectral classifications with NaCo photometry, we derived luminosities
and effective temperatures for these stars.  Due to the improved
magnitude limit of our sample relative to previous work, our derived
H-R diagram clearly shows the red clump, as well as the red giant
branch and asymptotic giant branch.  Using a maximum likelihood
analysis, we compared the observed distribution to models representing
ten possible star formation histories.  The best-fit model corresponds
to continuous star formation over the last 12 Gyr with a top-heavy
IMF.  The similarity of this result to the IMF observed for the most
recent epoch of star formation is intriguing and perhaps suggests a
connection between recent star formation and the stars formed
throughout the history of the Galactic Center.  The upcoming
FLAMINGOS-2 GC Survey on Gemini will provide important information
needed to further understand this suggestive result.

\acknowledgments
The authors thank Amiel Sternberg for his helpful
  discussions and insights.  H.M. is grateful to the Graduate Research
  Fellowship Program at NSF and the Graduate Opportunity Program
  Fellowship at UC Berkeley for funding this research.
  F.M. acknowleges support from the Alexander von Humboldt Foundation.
  T.A. is supported by Minerva grant 8563 and a New Faculty grant by
  Sir H. Djangoly, CBE, of London, UK.  This work has been supported
  by the National Science Foundation Science through AST 0205999 and
  the Center for Adaptive Optics, managed by the University of
  California at Santa Cruz under cooperative agreement No. AST
  9876783.  This work has made use of the IAC-STAR Synthetic CMD
  computation code. IAC-STAR is suported and maintained by the
  computer division of the Instituto de Astrofisica de Canarias.

{}

\end{document}